\documentclass[useAMS,usenatbib,usegraphicx]{mn2e}
\usepackage{amsmath,bm} 

\newcommand{\ob}{\overline}

\newcommand{\Rm}{R_{\rm m}}
\newcommand{\etat}{\eta_{\rm t}}
\title[Shear-driven and diffusive helicity fluxes]
{Shear-driven and diffusive helicity fluxes in $\alpha\Omega$ dynamos}

\author[G. Guerrero, P. Chatterjee and and A. Brandenburg]
{G. Guerrero$^{1}$\thanks{E-mail: 
guerrero@nordita.org (GG)} and
  P. Chatterjee$^{1}$
and A. Brandenburg$^{1,2}$
\\
$^{1}$Nordita, AlbaNova University Center, Roslagstullsbacken 23, SE
10691 Stockholm Sweden \\
$^{2}$Department of Astronomy, AlbaNova University Center, Stockholm
University, SE 10691 Stockholm, Sweden  
}
\begin{document}


\pagerange{\pageref{firstpage}--\pageref{lastpage}} \pubyear{2002}

\maketitle

\label{firstpage}

\begin{abstract}
We present nonlinear mean-field $\alpha\Omega$ dynamo simulations in
spherical geometry 
with simplified profiles of kinematic $\alpha$ effect and shear.
We take magnetic helicity evolution into account by solving
a dynamical equation for the magnetic $\alpha$ effect. This gives a  
consistent description of the quenching mechanism in mean-field dynamo 
models. The main goal of this work is to explore the effects of this
quenching mechanism in solar-like geometry, and in particular to
investigate the role of magnetic helicity fluxes, specifically
diffusive and Vishniac-Cho (VC) fluxes, at large magnetic Reynolds
numbers ($R_{\rm m}$). For models with  negative radial shear or
positive latitudinal shear, the magnetic $\alpha$ effect has
predominantly negative (positive) sign in the northern (southern)
hemisphere. In the absence of fluxes, we find that the
magnetic energy follows an 
$R_{\rm   m}^{-1}$ dependence, as found in previous works.
This catastrophic quenching is alleviated in models 
with diffusive magnetic helicity fluxes resulting in magnetic
fields comparable to the equipartition value even for
$R_{\rm m}=10^7$. On the other hand, models with a shear-driven
Vishniac-Cho flux show an increase of the amplitude of the magnetic
field with respect to models without fluxes, but only for 
$R_{\rm m}<10^4$.
This is mainly a consequence of assuming a vacuum outside the Sun
which cannot support a significant VC flux across the boundary.
However, in contrast with the diffusive flux, the VC flux
modifies the distribution of the magnetic field. 
In addition, if an ill-determined 
scaling factor in the expression for the VC flux is large enough,
subcritical dynamo action is possible that is driven by the action of shear
and the divergence of current helicity flux. 
\end{abstract}


\section{Introduction}
A crucial point in the study of astrophysical dynamos is to understand
the mechanism by which they saturate. Nevertheless, a consistent 
description of this process has rarely been considered in mean-field
dynamo (MFD) modeling and only a heuristic description is often used. 
An important phenomenon happens when the dynamo operates in closed or
periodic domains: the turbulent contribution to the dynamo
equation, i.e., the $\alpha$ effect, decreases for large values of the
magnetic Reynolds number. This process is known as catastrophic
quenching and can pose a problem in explaining the generation of
magnetic field in late type stars like the Sun or the Galaxy, where
$R_{\rm m}$ could be of the order of $10^9$ or $10^{15}$,
respectively.  

In the last few years the nature of the catastrophic quenching has
been identified as a consequence of magnetic helicity
conservation \citep[for a review see][]{bs05a}. It has been found
that in the nonlinear phase of the dynamo process, conservation of
magnetic helicity gives 
rise to a magnetic $\alpha$ effect ($\alpha_{\rm M}$) with a sign opposite 
to the inductive contribution due to the helical motions, i.e., the
kinematic $\alpha$ effect. As the production of $\alpha_{\rm M}$
depends on $R_{\rm m}$, the final value of the magnetic field should
also follow the same dependence. However, real astrophysical bodies
are not closed systems, but they have open boundaries that
may allow a flux of magnetic helicity. The shedding of magnetic
helicity may mitigate the  catastrophic $\alpha$ quenching. 

These ideas have been tested in direct numerical simulations (DNS) in 
both local Cartesian and global spherical domains. 
In the former \citep{B05,kapyla08} it has been clearly shown that open  
boundaries (e.g.\ vertical field boundary conditions)
lead to a faster saturation of a large-scale magnetic field compared
with cases in closed domains (perfect conductor or triple-periodic
boundary conditions). In the latter, it has been found that it is
possible to build up large-scale magnetic fields either with forced
turbulence \citep{B05,mitra10b} or with convectively driven turbulence
\citep[e.g.,][]{brownetal10,ketal10}.  These models generally used
vertical field boundary conditions. 

In flux-transport dynamos \citep{dc99,gdp08}
as well as in interface dynamos of the solar cycle
\citep[e.g.][]{mgc97a,cmg97b}
the quenching mechanism has been considered either through 
an {\it ad hoc} algebraic equation or by phenomenological
considerations \citep{cnc04}, but most of the time the models do not
consider the effects of magnetic helicity conservation. 
An exception is the recent paper by \cite{cbg10}, where these effects
have been considered in the context of an interface dynamo.

In general the magnetic helicity depends on time, 
so it is necessary to solve an additional dynamical equation for
the contribution of the small-scale field to  the magnetic helicity
together with the induction equation for the magnetic field. 
In the past few years, some effort has already been made to
consider  this dynamical saturation mechanism in MFD models like in
the 1D $\alpha^2$ dynamo models presented in \cite{bcc09}, in
axisymmetric models in cylindrical geometry for the 
galactic $\alpha \Omega$ dynamo \citep{setal06}, and also in models
with spherical geometry for an $\alpha^2$ dynamo \citep{betal07}.
The role of various kinds of magnetic
helicity fluxes have been explored in several papers
\citep{bcc09,zetal06,setal06}.

Our ultimate goal is to develop a self-consistent MFD model
of the solar dynamo, with observed velocity profiles and turbulent
dynamo coefficients computed from the DNS.
This is a task that
requires intensive efforts. Hence we shall proceed step by step,
starting with simple models and then including more realistic physics
on the way.  In this work we will study the effects of magnetic
helicity conservation in simplified $\alpha \Omega$ dynamo models for
a considerable number of cases. More importantly, we 
shall perform our calculations in spherical geometry, 
which is appropriate for describing stellar dynamos, with 
suitable boundary conditions, and considering 
shear profiles which are a simplified version of the observed solar
differential rotation. We shall also explore how magnetic helicity
fluxes affect the properties of the solution. Two classes of fluxes
are considered in this paper: a diffusive flux and a shear-driven or
Vishniac-Cho (hereafter VC) flux \citep{vch01}.
We consider models with either radial or latitudinal
shear. The effects of meridional circulation will be investigated in
detail in a companion paper \citep{cgb10}.

This paper is organized as follows: in Section~\ref{sec.aodyn} we
describe the basic mathematical formalism of the $\alpha\Omega$ dynamo, 
give the formulation of the equation for $\alpha_{\rm M}$ and also
justify the fluxes included. In Section~\ref{sec.model} we describe
the numerical method and then, we present our results in
Section~\ref{sec.results} starting from a dynamo model with algebraic
quenching to models with dynamical $\alpha$ quenching and different 
fluxes. Finally, we provide a summary of this work in
Section~\ref{sec.conc}.  

\section[]{The $\alpha \Omega$ dynamo model}
\label{sec.aodyn}

In mean-field dynamo theory, the evolution of the magnetic field is 
described by the mean-field induction equation,
\begin{equation}
  \label{eq1}
 \frac{\partial \ob{\bm{B}}}{\partial t} = \bm\nabla
 \times\left(\ob{\bm{U}} \times \bm{\ob{B}} + \bm{\ob{\cal{E}}} -
 \eta_{\rm m}\bm{\nabla}\times\bm{\ob{B}}\right), 
\end{equation}
\noindent
where $\bm{\ob{B}}$ and $\bm{\ob{U}}$ represent the mean magnetic
and velocity fields, respectively, $\eta_{\rm m}$ is the molecular 
diffusivity, $\bm{\ob{\cal{E}}}=\alpha \bm{\ob{B}} -
\etat\mu_0\bm{\ob{J}}$ is the mean electromotive force 
obtained using a closure theory like the first order smoothing 
approximation, where $\bm{\ob{\cal{E}}}$ gives the contribution of
the small-scale components on the large-scale field, $\alpha$ is the
non-diffusive contribution of the turbulence, $\etat$ is the turbulent
magnetic diffusivity, $\bm{\ob{J}}=\bm{\nabla}\times\bm{\ob{B}}/\mu_0$
is the mean current density, and $\mu_0$ is the vacuum permeability.

In spherical coordinates and under the assumption of axisymmetry,
it is possible to split the magnetic and the velocity
fields into their azimuthal and poloidal components, 
$\bm{\ob{B}}=B\bm{\hat{e}}_{\phi}+\bm{\nabla}\times
(A\bm{\hat{e}}_{\phi})$ and $\bm{\ob{U}}=r \sin\theta \Omega
\bm{\hat{e}}_{\phi} + \bm{u}_{\rm p}$, respectively. For the sake of
simplicity we shall not consider the meridional component of the flow,
i.e.\ ${\bm u}_{\rm p}=0$. Then, the 
toroidal and poloidal components of equation (\ref{eq1}) may be
written as 
\begin{eqnarray}
\frac{\partial B}{\partial t} &\!=\!& \label{eq.inb}
s {\bm B}_{\rm p} \cdot \bm\nabla \Omega
- \left[\bm\nabla \eta \times (\bm\nabla \times
B \bm{\hat{e}}_{\phi})\right]_{\phi} 
+ \eta D^2B\;,
\quad, \\ 
\frac{\partial A}{\partial t}
&\!=\!& \alpha B + \eta D^2A\;,
\label{eq.ina}
\end{eqnarray}
where $D^2=\nabla^2-s^{-2}$ is the diffusion operator, $\eta=\eta_{\rm
m}+\etat$, $s=r \sin \theta$ is the distance from the axis,
and $\bm{B}_{\rm p}=\bm\nabla\times(A\hat{e}_{\phi})$ is the poloidal field.

The two source terms in equations (\ref{eq.inb}) and (\ref{eq.ina}),
$s \bm{B}_{\rm p} \cdot \bm\nabla \Omega$ and $\alpha B$, express the
inductive effects of shear and turbulence,
respectively. The relative importance of these two  effects may be
quantified through the non-dimensional dynamo numbers: 
$C_{\Omega}=\Delta \Omega L^2 / \etat$ and
$C_{\alpha}=\alpha_0 L / \etat$, where $\Delta\Omega$ is the
angular velocity different between top and bottom of the domain.
Note that equations (\ref{eq.inb}) and (\ref{eq.ina}) are valid only
in the limit $C_{\Omega} \gg C_{\alpha}$, known as $\alpha\Omega$
dynamo. 

The inductive effects of the shear may be understood as the
stretching of the magnetic field lines due to the change in the
angular velocity between two adjacent points. 
On the other hand, the kinematic $\alpha$-effect is 
the consequence of helical motions of the plasma which produce
screw-like motions in the rising blobs of the magnetic field.
Using the first order smoothing approximation it
may be expressed as:
\begin{equation}
\alpha_{\rm K}=-\textstyle{\frac{1}{3}}\tau\overline{\bm{\omega}\cdot\bm{u}}\;,
\end{equation}
where, $\tau$ is the correlation time of the turbulent motions and
$\bm{\omega}=\bm{\nabla}\times\bm{u}$ is the small-scale vorticity.
The saturation value of the magnetic field may be obtained by
multiplying $\alpha_{\rm K}$ by the quenching function 
$f_{\rm q}=\left(1 + B^2/B_{\rm eq}^2 \right)^{-1}$, which saturates
the exponential growth of the magnetic field at values close to the  
equipartition field strength given by $B_{\rm eq}=(\mu_0
\overline{\rho{\bm u}^2})$.  
This form of algebraic quenching was introduced 
heuristically \cite[see, e.g.][]{stix72} and has been often used
as the standard quenching mechanism in many dynamo simulations.
However, it does not give information about the back reaction process 
and is independent of any parameter of the system like the magnetic
Reynolds number. A consistent description of the quenching mechanism
will be presented in the following section.

\subsection{Dynamical $\alpha$ effect}
\label{s.dyn.alpha}

Recently, it has been demonstrated that when the amplitude of the
magnetic field reaches values near the equipartition,
the $\alpha$-effect is modified by a magnetic
contribution, the so called magnetic $\alpha$ effect, denoted by
$\alpha_{\rm M}$. It is usually the case that $\alpha_{\rm M}$ has a
sign opposite to $\alpha_{\rm K}$ resulting thus in the saturation of
the magnetic field. \cite{pouquet76} have shown that $\alpha_{\rm M}$ is
proportional to the small-scale current helicity of the system, hence
it is possible to write $\alpha$ as a sum of two contributions, one
from the fluid turbulence and other from the magnetic field, as
follows: 
\begin{equation}
\alpha = \alpha_{\rm K} + \alpha_{\rm M} =
-{\textstyle\frac{1}{3}}\tau\overline{\bm{\omega}\cdot \bm{u}}
+\textstyle{\frac{1}{3}}\tau\overline{\bm{j}\cdot\bm{b}}/\overline{\rho} \;,
\label{eq.tot.alpha}
\end{equation}
\noindent
where $\overline{\rho}$ is the mean density of the medium, assumed here as
a constant, and $\bm{j}=\bm\nabla\times\bm{b}/\mu_0$ is the current density
of the fluctuating field. The mathematical expression that describes the
evolution of $\alpha_{\rm M}$ may be obtained by taking into account
the magnetic helicity evolution \citep{bb02}, which leads to:
\begin{equation}
\label{eq.am}
\frac{\partial \alpha_{\rm M}}{\partial t} = -2 \etat k_{\rm f}^2
\left(\frac{\bm{\overline{\cal{E}}}\cdot \overline{{\bm B}}}{B_{\rm eq}^2} 
 + \frac{\alpha_{\rm M}}{\Rm} \right) - \bm\nabla \cdot
 \bm{\overline{{\cal{F}}}}_{\alpha}\;,
\end{equation}
\noindent
where $k_{\rm f}=2 \pi / (L - r_{c})$ with $r_c=0.7L_0$ is a suitable
choice for the wave number of the forcing scale,
the magnetic Reynolds number $R_{\rm M}=\etat/\eta_{\rm m}$ and
$\bm{\overline{\cal{F}}}_{\alpha}$ is the flux of the magnetic $\alpha$
effect related to the flux of the small-scale magnetic helicity,
$\bm{\overline{F}}_{\rm f}$ through:  
\begin{equation} 
\label{eq.flux}
\bm{\overline{\cal F}}_{\alpha} = \frac{\mu_0 \overline{\rho} \etat
  k_{\rm f}^2}{B_{\rm eq}^2}\bm{\overline{F}}_{\rm f} \quad ,
\end{equation}
According to previous authors $\alpha_{\rm M}$ has a finite
value in the interior of the domain in absence of fluxes
($\bm{\overline{\cal{F}}}_{\alpha}=0$), and its 
sign is usually opposite to the sign of $\alpha_{\rm K}$ in such a
way that the final amplitude of the total $\alpha$-effect decreases,
and so does the final value of the magnetic energy. 

\subsection{Magnetic helicity fluxes}

Recently it has been pointed out that the catastrophic quenching could 
be alleviated by allowing the flux of small-scale magnetic (or
current) helicity out of the domain, so that the total 
magnetic helicity inside need not be conserved any longer. 
Alternately, we may introduce those fluxes in the equation for
$\alpha_{\rm M}$; see equation (\ref{eq.flux}).
Several candidates have been proposed for the helicity fluxes in the 
past \citep{kr99,vch01,sb04}. Amongst them are the flux of
magnetic helicity across the iso-rotation contours,
advective and diffusive fluxes and also the explicit removal of
magnetic helicity in processes like coronal mass ejections or
galactic fountain flows, for the case of the galactic dynamo.

From the mathematical point of view, the nature of the flux terms in
the equation for $\alpha_{\rm M}$ has not been demonstrated with
sufficient rigor. 
However, several DNS have pointed to its existence.  

Firstly, the shearing box convection simulations of
\cite{kapyla08} showed that in the presence of open boundaries,
the large-scale magnetic field grows on temporal scales 
much shorter than the dissipative time scale. They concluded from 
this that open boundaries may allow the magnetic helicity to escape  
out of the system. These experiments seem to be compatible with the
flux proposed by 
\cite{vch01}, whose functional form may be expressed as 
\cite[see][for further details]{sb04,bs05c}: 
\begin{equation}
\label{eq.vcf}
\overline{\cal F}^{\rm VC}_i
= C_{\rm VC} \epsilon_{ijl} \overline{\mathsf{S}}_{lk}
\overline{B}_j\overline{B}_k \quad,
\end{equation}
\noindent
where $\overline{\mathsf{S}}_{lk}=\frac{1}{2}(\overline{U}_{l,k} +
\overline{U}_{k,l})$ is the mean rate of strain tensor and $C_{\rm VC}$ is
a non-dimensional scaling factor. 
As we assume ${\bm u}_{\rm p}=0$,
this flux has the following three components: 
\begin{eqnarray}
\overline{\cal F}^{\rm VC}_r &=& C_{\rm VC} \label{eq.vcfr}
\left[\overline{\mathsf{S}}_{\phi r}  B_{\theta}B_{\rm r} 
+  \overline{\mathsf{S}}_{\theta \phi} (B_{\theta}^2 - B_{\phi}^2) \right]  \;,
\\
\overline{\cal F}^{\rm VC}_{\theta} &=& C_{\rm VC} \label{eq.vcft}
\left[- \overline{\mathsf{S}}_{\phi \theta}  B_{r}B_{\theta}  
+  \overline{\mathsf{S}}_{r \phi} (B_{\phi}^2 - B_r^2) \right]  \;,
\\
\overline{\cal F}^{\rm VC}_{\phi} &=& C_{\rm VC} \label{eq.vcfp}
\left[\overline{\mathsf{S}}_{\theta \phi}  B_{r}B_{\phi} 
-  \overline{\mathsf{S}}_{r \phi} (B_{\theta}^2 - B_{\phi}^2) \right] 
\;,
\end{eqnarray}
\noindent
with $\overline{\mathsf{S}}_{\phi r} = \overline{\mathsf{S}}_{r \phi}
= r \sin\theta (\partial \Omega / \partial r)/2$ and
$\overline{\mathsf{S}}_{\theta \phi} = \overline{\mathsf{S}}_{\phi
  \theta} = \sin\theta (\partial \Omega / \partial \theta)/2$.  

Secondly, \cite{mitra09}  performed $\alpha^2$ dynamo simulations 
driven by forced turbulence in a box with an equator.
They found that the diffusive flux of $\alpha_{\rm M}$  across the
equator can be fitted to a Fickian diffusion law given by,
\begin{equation}
\mbox{\boldmath ${\cal F}$}_{\rm D}
= -\kappa_{\alpha}(r) \bm\nabla \alpha_{\rm M} \; .
\label{eq.fickian}
\end{equation}
\noindent
They also computed the numerical value of this diffusion coefficient,
and found it to be of the order of turbulent diffusion coefficient. 
They also found that the time averaged flux is gauge independent. 
Both results were later corroborated by simulations without equator,
but with a decline of kinetic helicity toward the boundaries
\citep{HB10}. 

Additionally, magnetic helicity may be advected by the mean velocity 
with a flux given by $\bm{\overline{\cal F}}_{\rm ad}=\alpha_{\rm M} \overline{\bm
  U} $, or it may be expelled from the solar interior by 
coronal mass ejections (CMEs) or by the solar wind. This flux, ${\cal
  F}_{\rm CME}$, may account for $\sim10$\% of the total helicity generated by
the solar differential rotation, as estimated by \cite{br00}. It can
be modeled by artificially removing a small amount of $\alpha_{\rm M}$ 
every $\tau$ time \citep{bcc09}, or also by a radial velocity field
that mimics the solar wind.

The total flux of magnetic helicity may be written as the sum of these 
contributions,
\begin{equation}
\bm{\overline{\cal F}} = \bm{\overline{\cal F}_{\rm VC}} + \bm{\overline{\cal F}}_{\rm D}
+ \bm{\overline{\cal F}}_{\rm ad} + \bm{\overline{\cal F}}_{\rm CME} \; . 
\label{eq.fluxes}
\end{equation}
Since in this dynamo model we do not include any component of
the velocity field other than the differential rotation,
in this study we will consider only the first two terms on the 
rhs of equation (\ref{eq.fluxes}).

\section{The model}
\label{sec.model}

\begin{figure}
\includegraphics[width=\columnwidth]{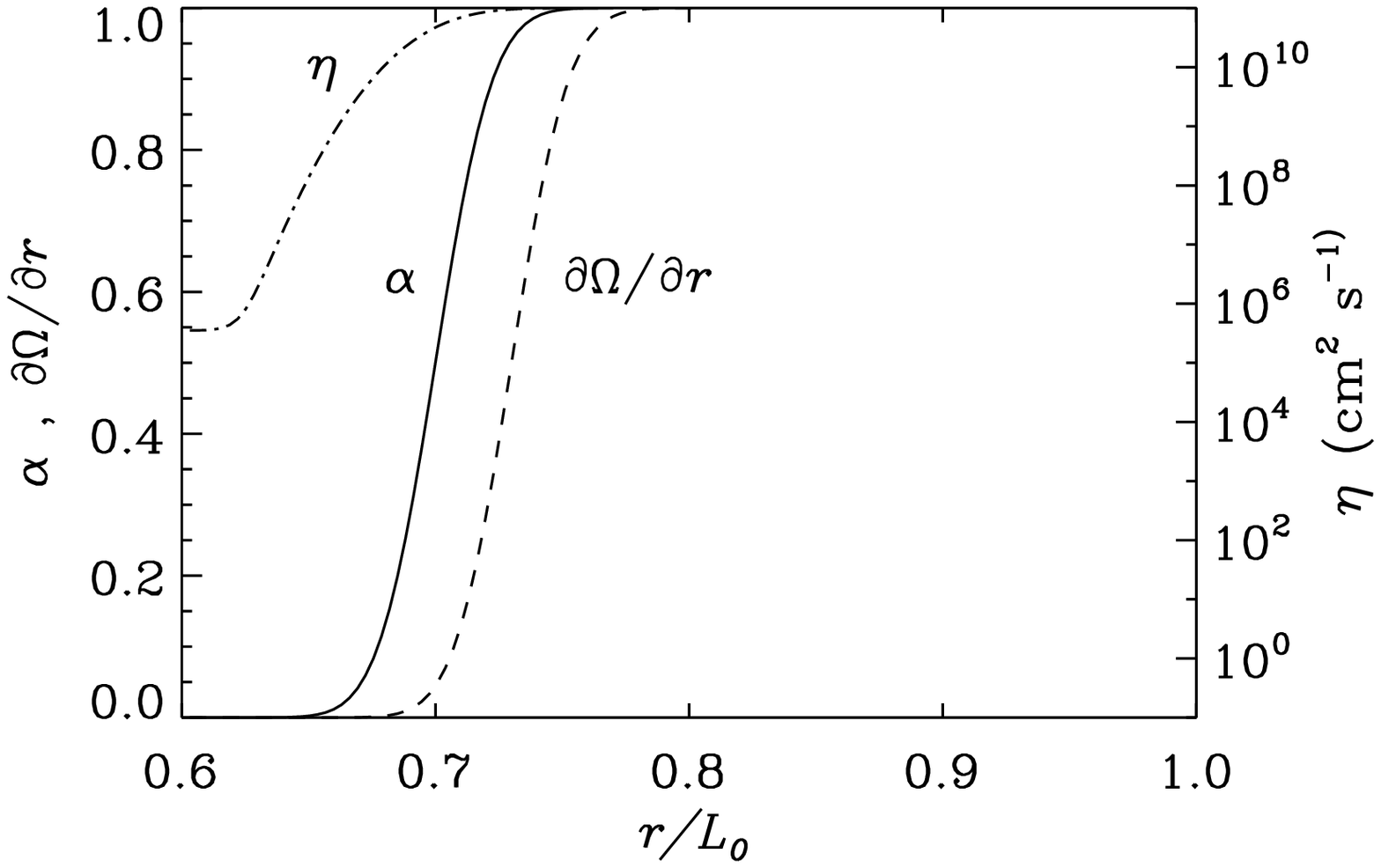}\\
\includegraphics[width=\columnwidth]{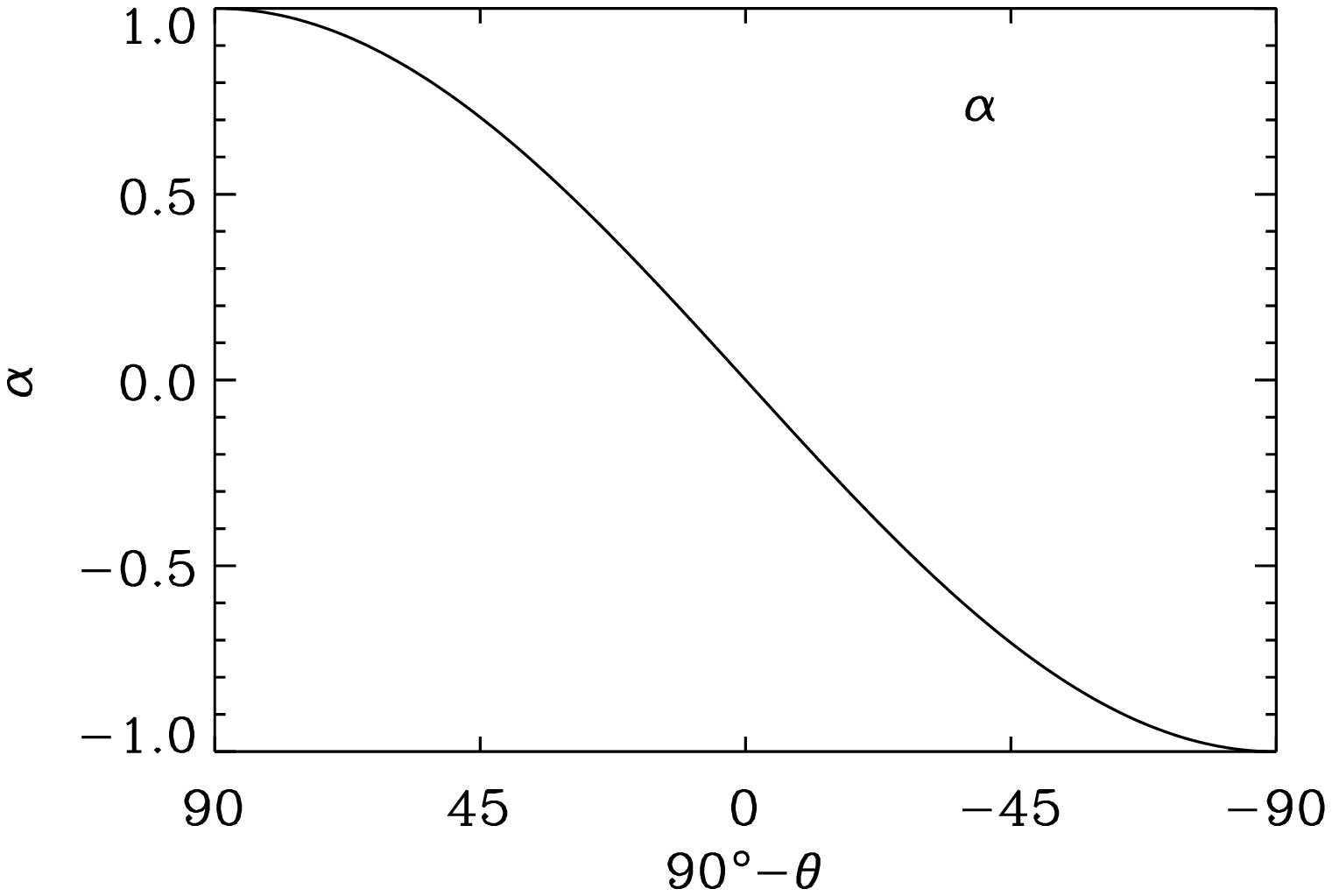}
\caption{Profiles of the dynamo ingredients, $\alpha_{\rm K}$ (solid line),
  $\partial \Omega / \partial r$ (dashed line)  and $\etat$
  (dot-dashed line). All the values are normalized to their maximum
  value.}
\label{fig.profiles}
\end{figure}

We solve equation (\ref{eq.inb}), (\ref{eq.ina}) and (\ref{eq.am}) 
for $A$, $B$ and $\alpha_{\rm M}$ 
in the meridional plane in the range
$0.6L\le r\le L$ and $0\le\theta\le\pi$.
We consider two different layers inside the spherical
shell. In the inner one the dynamo production terms are zero and 
go smoothly to a finite value in the external layer. 
The magnetic diffusivity changes from a molecular to a turbulent
value from the bottom to the top of the domain. This is achieved 
by considering error function profiles for the
magnetic diffusivity, the differential rotation, and the kinetic
$\alpha$ effect, respectively (see Fig.~\ref{fig.profiles}):
\begin{eqnarray}
\label{eq.eta}
\eta(r) &=& \eta_{\rm m} + \etat \Theta(r,r_1,w_1) \;, \\
\label{eq.omega}
\frac{\partial \Omega}{\partial r}(r) &=& C_{\Omega}
\left(\frac{L^2}{\etat} \right) \Theta(r,r_2,w_1) \;,\\
\label{eq.alpha}
\alpha_{\rm K}(r,\theta) &=& C_{\alpha}\left(\frac{L}{\etat} \right)
\Theta(r,r_1,w_1) \cos\theta \;,
\end{eqnarray}
\noindent
where $\Theta(r,r_{1,2},w)= \frac{1}{2}\left[1+ \mathrm{erf}
  \left\lbrace(r - r_{1,2})/{w_1} \right\rbrace \right]$, with  $r_1=0.7L_0$,
$r_2=0.72L_0$ and $w_1=0.025L_0$. We fix $C_{\Omega}=-10^{4}$ and vary
$C_{\alpha}$. 

The boundary conditions are chosen as follows: at the poles,
$\theta=0,\pi$, we impose $A=B=0$; at the
base of the domain, we impose a perfect conductor boundary condition,
i.e.\ $A=\partial(r B)/\partial r=0$. 
Unless noted otherwise, we use at the top a vacuum condition by coupling
the magnetic field inside with an external potential field, i.e.,
$(\nabla^2 - s^{-2})A=0$. A good description of the numerical
implementation of this boundary condition may be found in
\cite{dc94}. 

The equations for $A$ and $B$ are solved using a second-order
Lax-Wendroff scheme for the first derivatives, and centered finite
differences for the second-order derivatives. The temporal
evolution is computed by using a 
modified version of the ADI method of \cite{pr55} as explained in
\cite{dc99}. This numerical scheme has been used previously in several 
works on the flux-transport dynamo and the results were found to be in
good agreement with those using other numerical techniques
\citep{gdp07,gdp08,gddp09}.  

In the absence of magnetic helicity fluxes, equation (\ref{eq.am})
for $\alpha_{\rm M}$ corresponds to an initial value problem that can be
computed explicitly.  
However, as we are going to include a diffusive flux, we use for
$\alpha_{\rm M}$ the same numerical technique used for $A$ and $B$. 
All the source terms on the right hand side of equation (\ref{eq.am}) are 
computed explicitly. 
We have tested the convergence of the solution 
for $64^2$, $128^2$, and $256^2$ grid points. For cases with
small $\Rm$, there are no significant differences between different
resolutions, but for high $\Rm$, $64^2$ grid points is insufficient to
properly resolve the sharp diffusivity gradient. A resolution of
$128^2$ grid points is a good compromise between accuracy and speed.  

\section{Results}
\label{sec.results}

\subsection{ $\alpha \Omega$ dynamos with algebraic quenching}
\label{sec.4.1}

\begin{table*}
 \caption{Summary of main parameters and results of the numerical
   simulations.} 
 \label{table1}
 \begin{tabular}{@{}lccccccc}
  \hline
  Run   & $C_{\alpha}$ &  $R_{\rm m}$ & $\kappa_{\alpha}$  ($\etat$) & $C_{\rm VC}$ 
        & $\overline{\bm B}_{\rm rms}/B_{\rm eq}$ 
        & $T (L_0^2 / \etat)$ & $t$ ($L_0^2 / \etat$) \\
  \hline
  Ca$^{\rm C}$ & 1.975 & $10$  & - & - & 0.0008 & 0.0486 & $1.0$ \\
  Ca2.0 & 2.0 & $10$  & - & - & 0.15 & 0.0484 & $1.0$  \\
  Ca2.1 & 2.1 & $10$  & - & - & 0.33 & 0.0477 & $3.0$ \\
  Ca2.2 & 2.2 & $10$  & - & - & 0.45 & 0.0471 & $3.0$ \\
  Ca2.3 & 2.3 & $10$  & - & - & 0.56 & 0.0464 & $3.0$ \\
  Ca2.4 & 2.4 & $10$  & - & - & 0.64 & 0.0460 & $3.0$ \\
  Ca2.5 & 2.5 & $10$  & - & - & 0.69 & 0.0455 & $3.0$ \\
  \hline 
  Rm10 & 2.5 & $10$  & - & - & 0.21 & 0.0422 & $4.0$  \\
  Rm50 & 2.5 & $50$  & - & - & 0.25 & 0.0446 & $4.0$  \\
  Rm1e2 & 2.5 & $100$  & - & - & 0.2 & 0.0455 & $4.0$ \\
  Rm1e3 & 2.5 & $10^3$  & - & - & 0.07 & 0.0464 & $4.0$ \\
  Rm2e3 & 2.5 & $2$$\times$$10^3$  & - & - & 0.05 & 0.0464 & $6.0$ \\
  Rm5e3 & 2.5 & $5$$\times$$10^3$  & - & - & 0.03 & 0.0468 & $15.0$ \\
  Rm1e4 & 2.5 & $10^4$  & - & - & 0.02 & 0.048 & $15.0$ \\
\hline 
  DRm10 & 2.5 & $10$  & $0.005$    & - & 0.22 & 0.0422 & $4.0$ \\
  DRm50 & 2.5 & $50$  & $0.005$    & - & 0.26 & 0.0446 & $4.0$ \\
  DRm1e2 & 2.5 & $100$  & $0.005$  & - & 0.20 & 0.0455 & $4.0$ \\
  DRm1e3 & 2.5 & $10^3$  & $0.005$ & - & 0.09 & 0.0460 & $4.0$ \\
  DRm1e4 & 2.5 & $10^4$  & $0.005$ & - & 0.06 & 0.0457 & $5.0$ \\
  DRm1e5 & 2.5 & $10^5$  & $0.005$ & - & 0.05 & 0.0460 & $7.0$ \\
  DRm1e6 & 2.5 & $10^6$  & $0.005$ & - & 0.05 & 0.0457 & $8.0$ \\
  DRm1e7a & 2.5 & $10^7$  & $0.001$ & - & 0.026 & 0.0457 & $20.0$ \\
  DRm1e7b & 2.5 & $10^7$  & $0.005$ & - & 0.05 & 0.0460  & $10.0$ \\
  DRm1e7c & 2.5 & $10^7$  & $0.01$ & - & 0.073 & 0.0460 & $10.0$ \\
  DRm1e7d & 2.5 & $10^7$  & $0.03$ & - & 0.12 & 0.0460 & $8.0$\\
  DRm1e7e & 2.5 & $10^7$  & $0.05$ & - & 0.15 & 0.0457 & $4.0$\\
  DRm1e7f & 2.5 & $10^7$  & $0.1$ & - & 0.20 & 0.0460 & $4.0$\\
  DRm1e7g & 2.5 & $10^7$  & $1.0$ & - & 0.54 & 0.0458 & $4.0$\\
  DRm1e7h & 2.5 & $10^7$  & $5.0$  & - & 1.23 & 0.060 & $4.0$\\
  DRm1e7i & 2.5 & $10^7$  & $10.0$ & - & 1.76 & 0.0457 & $4.0$\\
\hline
  VCa & 2.5 & $10^3$  & - & 0.002 & 0.032 & 0.0449 & $4.0$\\
  VCb & 2.5 & $10^3$  & - & 0.01 & 0.02 & 0.0442 & $4.0$\\
  VCc & 2.5 & $10^3$  & - & -0.002 & 0.02 & 0.0447 & $4.0$\\
  VCd & 2.5 & $10^4$  & - & -0.002 & - & - & $4.0$\\
  VCD & 2.5 & $10^3$  & $0.1$ & 0.001 & 0.11 & 0.0446 & $4.0$\\
\hline
  Re1e3$_{\theta}$ & 2.5 & $10^3$  & - & - & 0.023 & 0.0282 &
  $8.0$\\ 
  VC$_{\theta}$a & 2.5 & $10^3$  & - & 0.004 & 0.04 & 0.033 &
  $4.0$\\ 
  VCD$_{\theta}$ & 2.5 & $10^3$  & 0.1 & 0.004 & 0.062 & 0.0266 &
  $4.0$\\ 
  Re1e3$_{\theta}$vf & 2.5 & $10^3$  & - & - & 0.036 & 0.032 &
  $6.0$\\ 
  VC$_{\theta}$vf & 2.5 & $10^3$  & - & 0.004 & 0.075 & 0.033 &
  $6.0$\\ \hline
 \end{tabular}
\end{table*}

\begin{figure}
  \includegraphics[width=84mm]{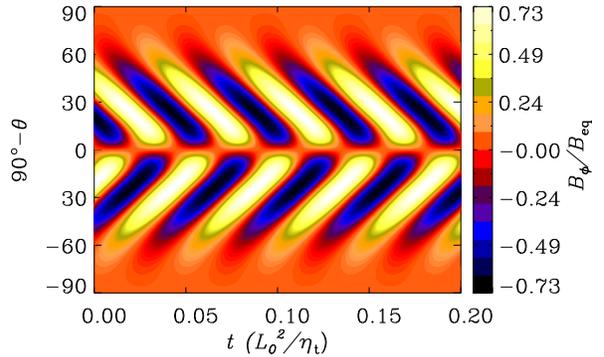}
\caption{Time-latitude butterfly diagram for the toroidal component of
  the magnetic field at $r=0.72L_0$, for an $\alpha \Omega$ dynamo 
  model with $C_{\Omega}=-10^4$ and $C_{\alpha}=2.5$.}
\label{fig.fig2}
\end{figure}

\begin{figure}
\includegraphics[width=80mm]{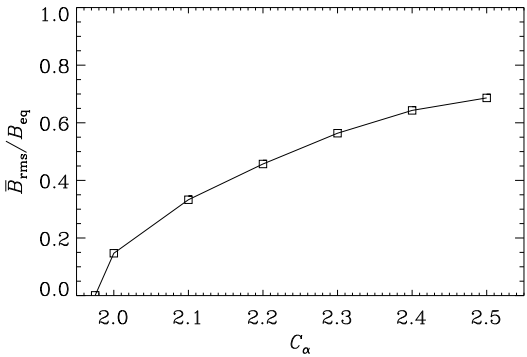}
\caption{Magnetic field average amplitude as a function of
  $C_{\alpha}$ using an algebraic quenching function
  that is independent of $R_{\rm m}$.}
\label{fig.bif}
\end{figure}

In order to characterize our $\alpha\Omega$ dynamo model we start by
exploring the properties of the system when the saturation is
controlled by algebraic quenching with $f_q=(1 + B^2/B_{\rm eq}^2)^{-1}$. 
We found that, with the profiles given by
equations (\ref{eq.eta})--(\ref{eq.alpha}), Fig.~\ref{fig.profiles},
the critical dynamo number is around $2 \times 10^4$ (i.e.,
$C_{\alpha}^{\rm C}=1.975$). The solution for the model is a dynamo wave
traveling towards the equator since it obeys the Parker-Yoshimura sign 
rule (see Fig.~\ref{fig.fig2}). In this case, the maximum amplitude of
the magnetic field depends only on the dynamo number of the system,
$C_{\alpha} C_{\Omega}$, as can be seen in the 
bifurcation diagram in Fig.~\ref{fig.bif}.
The quenching formula is here independent of $\Rm$, so the saturation
amplitude is also independent on $\Rm$.

\subsection{$\alpha \Omega$ dynamos with dynamical quenching}

In this section we consider dynamo saturation through the
dynamical equation for $\alpha_{\rm M}$ described in
Section~\ref{s.dyn.alpha}. In this models we distinguish three different
stages in the time evolution of the magnetic field: a growing phase, a
saturation phase and a final relaxation stage (see panels a, b and c
of Fig.~\ref{snaps.no.flux}).
The magnetic field is
amplified from its initial value, $5 \times 10^{-4} B_{\rm eq}$,
following an exponential growth. From the earliest stages of the
evolution we notice the growth of $\alpha_{\rm M}$ with
values that are predominantly negative in the northern hemisphere and
positive in the southern hemisphere. The latitudinal distribution of
$\alpha_{\rm M}$ is fairly uniform in the active dynamo region,
spanning from the equator to $\sim 60^\circ$ latitude.
The radial distribution exhibits two narrow layers where the sign of
$\alpha_{\rm M}$ is opposite to the dominant one developing at 
each hemisphere.  These are located at the base of the dynamo region   
($r\sim0.7 L_0$) and at a thin layer near to the surface
($r>0.95$). In the equation for the magnetic $\alpha$
effect, equation (\ref{eq.am}), the production term is proportional to 
$\bm{\ob{\cal{E}}} \cdot \overline{\bm{B}}=
\alpha \overline{\bm{B}}^2 - \etat\mu_0\bm{\ob{J}} \cdot \ob{\bm{B}}$.
The first component of
this term has the same sign as $\alpha_{\rm K}$, which in general is
positive in the northern and negative in the southern
part of the domain. The minus sign in front of the right hand side
of equation (\ref{eq.am}) defines then the sign of $\alpha_{\rm M}$.
However, at the base and at the top of the dynamo region, 
$\alpha_{\rm K}\rightarrow 0$ and $B\rightarrow 0$, respectively.
The term $\etat \bm{\ob{J}} \cdot \bm{\ob{B}}$ is the only source of
$\alpha_{\rm M}$ and leads to the formation of these two thin layers. 

The space-time evolution of $\alpha_{\rm M}$ depends on the value of
the magnetic Reynolds number. For small $R_{\rm m}$, the decay 
term in equation \ref{eq.am} (i.e.\ the second term in the
parenthesis) becomes 
important, so that there is a competition between the production and 
decay terms resulting in an oscillatory behavior in the
amplitude of the magnetic $\alpha$ effect, as is indicated by the 
vertical bars in the middle panel of Fig.~\ref{fig.brms}. 
The period of these oscillations is the half the period of the
magnetic cycle.  With increasing $\Rm$, 
the amplitude of the oscillations decreases such that for
$R_{\rm m}\le10^3$, $\alpha_{\rm M}$ is almost steady.

The morphology of the magnetic field corresponds to a multi-lobed
pattern of alternating polarity (left panels of
Fig. \ref{snaps.no.flux}). These lobes are radially distributed 
in the whole dynamo region with maximum amplitude at the base of this
layer. The poloidal magnetic field follows a similar pattern with
lines that are open at the top of the domain due to the potential
field boundary condition. There is a phase shift between
toroidal and poloidal components which we have estimated to be
$\sim0.4 \pi$. The model preserves the initial dipolar parity during 
the entire evolution.

\begin{figure}
\includegraphics[width=84mm]{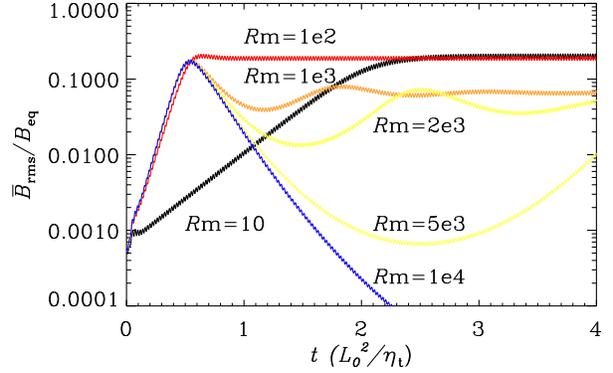}
\caption{Time evolution of the averaged mean magnetic field for
  different values of $\Rm$. Note that for $\Rm>10^3$, we have
  allowed the simulations to evolve more than $4$ diffusion times, as
  indicated in Table \ref{table1}.}
\label{fig.no.flux}
\end{figure}

\begin{figure}
\includegraphics[width=\columnwidth]{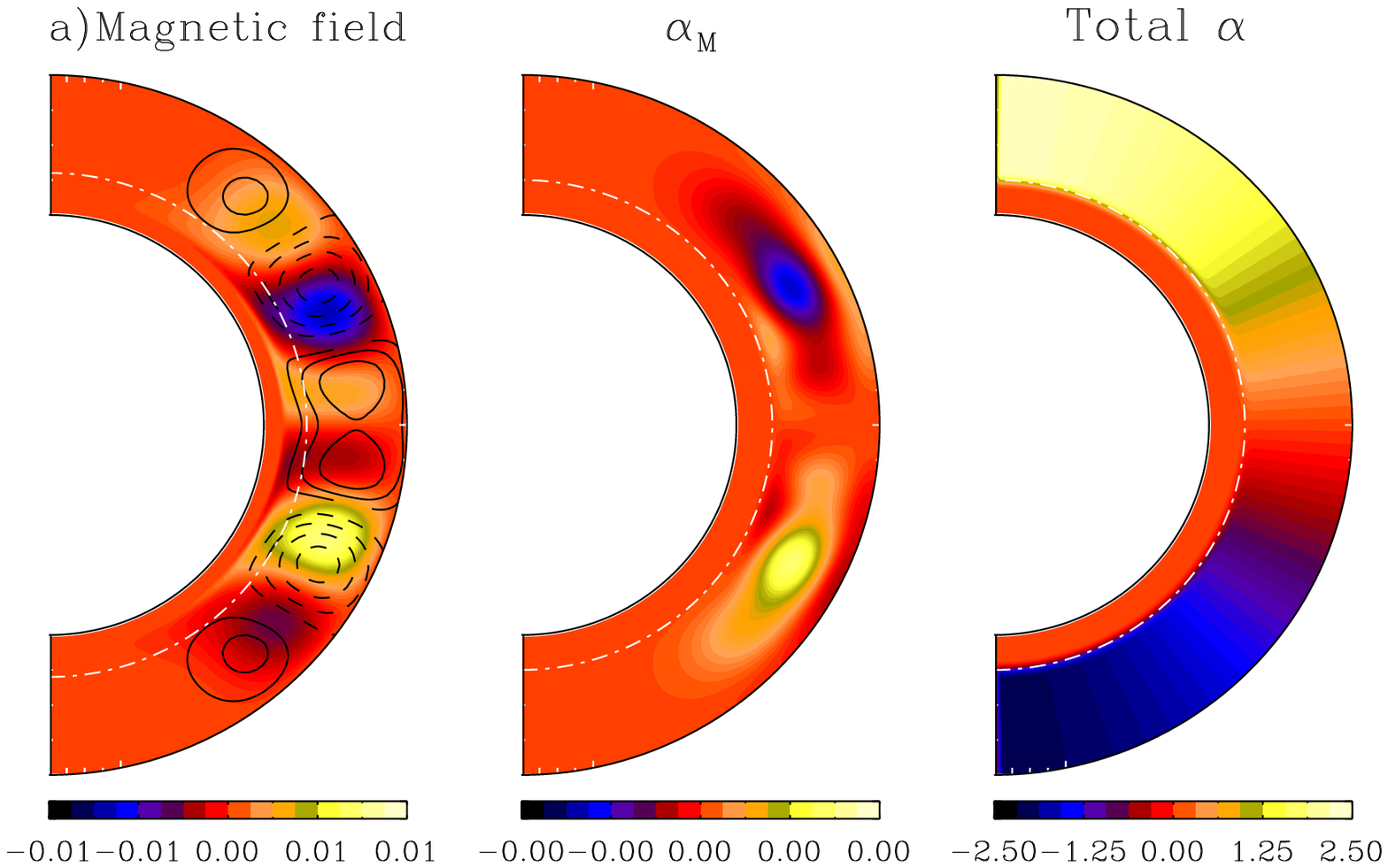}\\
\includegraphics[width=\columnwidth]{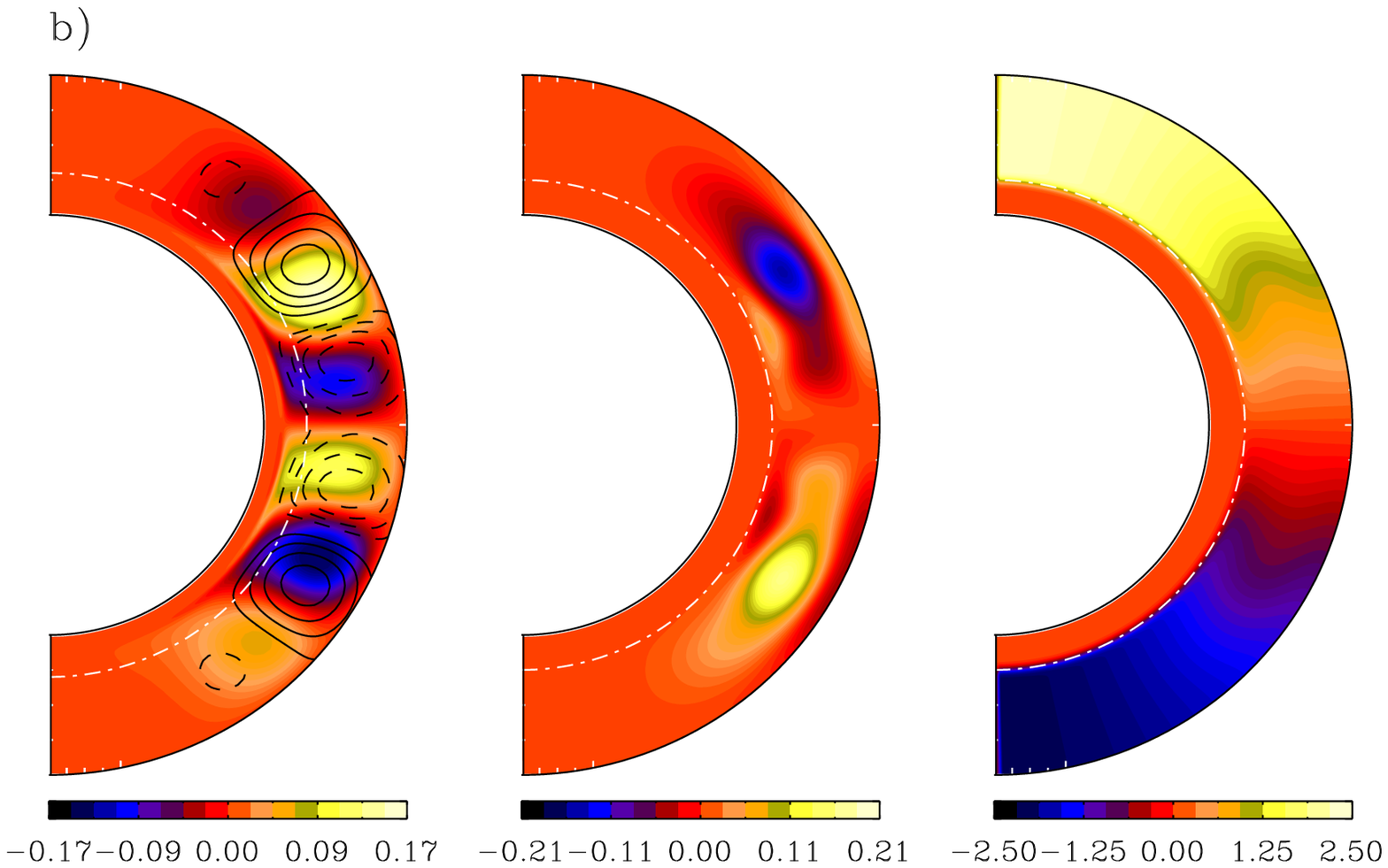}\\
\includegraphics[width=\columnwidth]{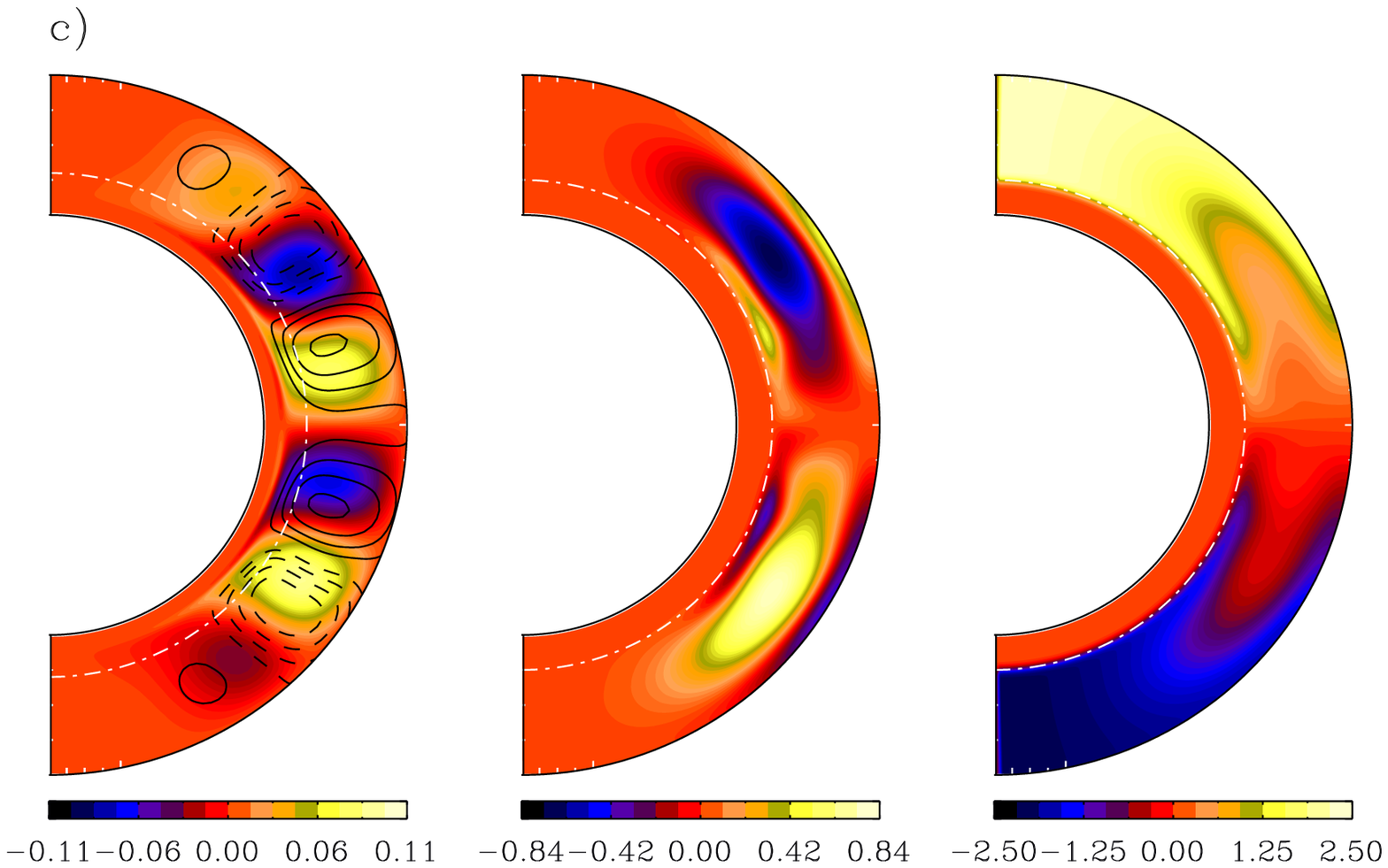}\\
\caption{Meridional snapshots of three different phases of evolution
  of the dynamo model with dynamical quenching, a)
  $t=0.25$ ($L_0^2/\etat$), b) $t=0.5$ ($L_0^2/\etat$)
  and c) $t=2.0$ ($L_0^2/\etat$). The left panel shows the
  contours of toroidal magnetic field in color scale, and positive
  (negative) poloidal magnetic field lines in continuous (dashed)
  lines. The central panel shows the distribution of $\alpha_{\rm M}$, and
  the right panel shows the distribution of the residual $\alpha$. All
  values are in non-dimensional units (i.e., $\overline{\bm
    B}/B_{\rm eq}$), so that the color scale is different for each 
  figure as indicated in the respective color bar.}
\label{snaps.no.flux}
\end{figure}

The evolution of $\alpha_{\rm M}$ traces the growth of the 
magnetic field, but its final value depends on the magnetic Reynolds
number. For small $\Rm$, after saturation, $\alpha_{\rm M}$ reaches a
steady state, but for large $\Rm$, its relaxation is modulated by
over-damped oscillations. The relaxation time is proportional to
$\Rm$, which means that for $\Rm\gg1$ the simulation must run
for many diffusion times. 
The differences in the relaxation time observed for $\alpha_{\rm M}$ 
reflects the evolution of the magnetic field, as is shown in
Fig.~\ref{fig.no.flux}.

We observe that the rms  value of the magnetic field remains steady 
during the saturation phase for $\Rm<10^2$.
For $10^2<\Rm<10^3$, a bump appears in the curve of
magnetic field evolution, followed by the relaxation to a steady
value, whereas for $\Rm>10^3$,  the magnetic energy shows
over-damped relaxations with a final energy proportional to
$\rm{Rm}^{-1}$ as has been previously reported \citep{betal07}.  These
oscillations in the time evolution plot of the averaged magnetic
field have been reported in mean 
field dynamo simulations including the dynamical $\alpha$-effect
\citep{bs05c}. 

Not many DNS of $\alpha\Omega$
dynamo exist so far in the literature with $\Rm \ge 100$ in order to
compare with our results. However, in the local $\alpha\Omega$  
dynamo simulations of \cite{kapyla08}, a rapid decay of the
magnetic field seems to occur after the initial saturation for
moderate values of 
$\Rm$. This decay forms a bump in the curve of the averaged
magnetic field (see their Fig. 14), similar to the bump that we
obtain for $10^2<\Rm<10^3$. 

For reasons of clarity in the
Fig. \ref{fig.no.flux} we do not show the 
entire time evolution of each simulation with  $\Rm>10^3$. The
total evolution time as well as the final value of the magnetic 
field of each simulation are shown in the Table \ref{table1}. 
For magnetic Reynolds numbers above $2\times10^4$, the initial
kinematic phase is followed by a decay phase during which the total
$\alpha$ effect goes through subcritical values and then the dynamo 
fails to start again.

In Fig. \ref{snaps.no.flux} we present the meridional distribution of
the magnetic field (left panel), $\alpha_{\rm M}$ (middle
panel) and the total $\alpha$ (right panel), in normalized
units, for the three different stages of evolution corresponding
to the early kinematic phase, the late kinematic phase and the
saturated phase. 
These snapshots correspond to the simulation with $\Rm=10^3$
(Run~Rm1e3 in Table \ref{table1}).  
The multi-lobed pattern of the toroidal field 
represented with filled contours remains unchanged during the
evolution even though its amplitude increases.  
The same occurs for the poloidal component, shown by continuous and
dashed streamlines for positive and negative values, respectively.  

The magnetic $\alpha$ effect (middle panels) is formed first at
latitudes between $\pm30^{\circ}$ and then it amplifies and expands
to latitudes up to $\sim \pm60^{\circ}$. This makes the total
$\alpha$ effect, initially similar to $\alpha_{\rm K}$
(Fig. \ref{fig.profiles} and top panel of Fig. \ref{snaps.no.flux}a),
smaller at lower 
latitudes in the central area of the dynamo region. At the bottom and
at the top of the domain $\alpha_{\rm M}$ and $\alpha_{\rm K}$ have
the same sign making the total $\alpha$ larger. However, the
global effect is a decrease of the dynamo efficiency. 

\subsection{Diffusive flux for $\alpha_{\rm M}$}
\begin{figure}
\includegraphics[width=\columnwidth]{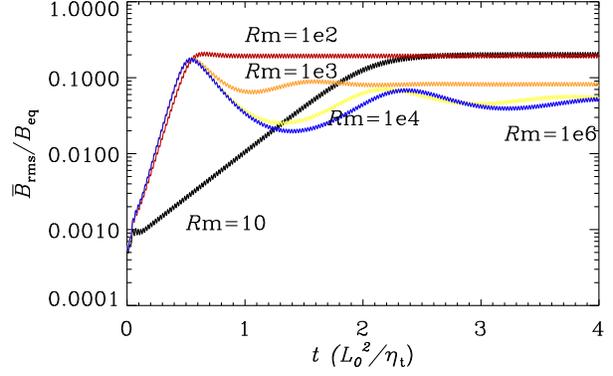}
\caption{The same that Fig. \ref{fig.no.flux} but for simulations
  including a diffusive flux of $\alpha_{\rm M}$. All the simulations
  correspond to $\kappa_{\alpha}=0.005 \etat$.}
\label{fig.dif.flux}
\end{figure}

\begin{figure}
\includegraphics[width=84mm]{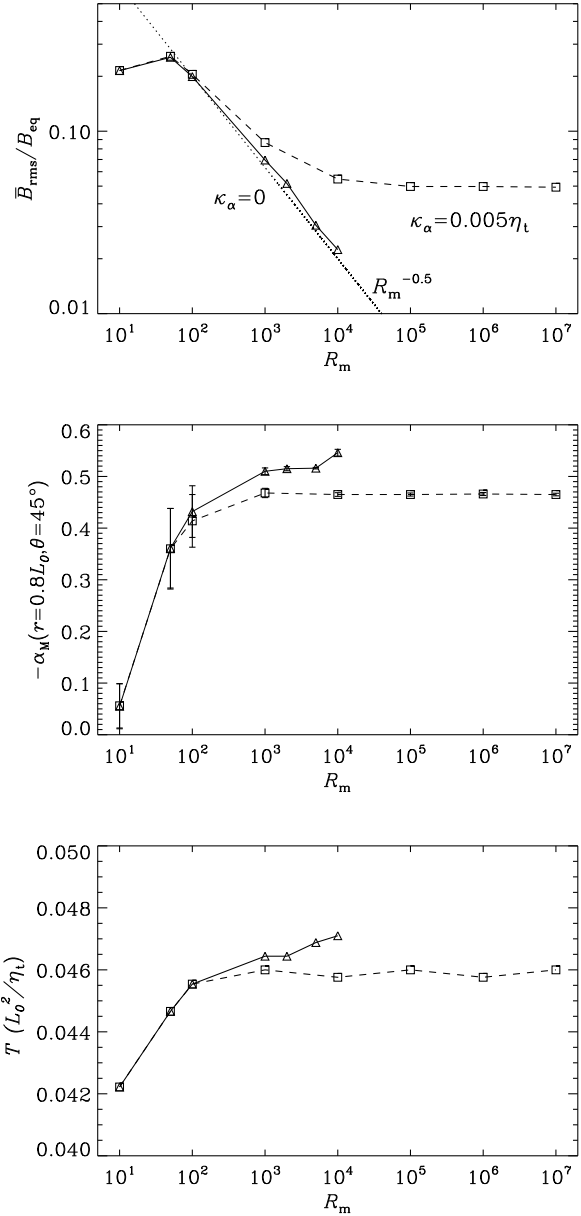}
\caption{$R_{\rm m}$ dependence of the averaged mean magnetic field
  (top), the temporal mean value of ${\alpha_{\rm M}}$ at
  $r=0.8L_0$, $\theta=45^\circ$ (middle) and the dynamo
  cycle period, $T$ in diffusion time units (bottom). The continuous
  line present the result for simulations without $\alpha_{\rm M}$
  diffusive flux ($\kappa_{\alpha}=0$) and the dashed line shows
  the results for  $\kappa_{\alpha}=0.005\etat$. The error lines
  in the middle panel indicate the maximum and minimum amplitudes in
  the oscillations of $\alpha_{\rm M}$ at that point.}
\label{fig.brms}
\end{figure}

In this section we consider a Fickian diffusion term in equation 
(\ref{eq.fickian}) for $\alpha_{\rm M}$.
We consider a diffusion coefficient varying
from  $5\times 10^{-3} \etat$ to $10\,\etat$ in the dynamo region and
with $\kappa_{\alpha}=\eta_{\rm m}$ in the bottom layer. In these
cases, the initial evolution of  $\alpha_{\rm M}$ is similar 
to the cases presented in the previous section: negative (positive)
values for $\alpha_{\rm M}$ in the northern (southern) hemisphere, with
narrow regions of opposite values nearby the regions where $\alpha_{\rm
  K}=0$ or ${\bm B}=0$. However, at the later stages,
$\alpha_{\rm M}$ is much more diffuse in the 
entire domain and has only one sign in each hemisphere. 
This is the result of cancellation of $\alpha_{\rm M}$ with opposite
signs occurring in each hemisphere due to radial diffusion.
Contrary to the cases without fluxes, we now obtain finite values of
$B_{\rm sat}$ for large values of $\Rm$, as can be seen in
Fig. \ref{fig.dif.flux}. All the cases depicted in this figure
correspond to $\kappa_{\alpha}=0.005 \etat$. We notice that the
final value of the magnetic field still remains small compared to the
equipartition ($\le 0.1 B_{\rm eq}$), but it is clear that even this very
modest diffusion prevents the $\alpha$ effect from being
catastrophically quenched. This is also evident from the top panel of
Fig.\ \ref{fig.brms}, where we plot the final strength of 
$\ob{\bm B}$ as a function of $\Rm$, for the cases with and
without dissipative flux. In the middle and bottom panels of the
Fig.\ \ref{fig.brms} we compare the behavior of the normalized
${\alpha_{\rm M}}$, at a given point inside the dynamo region, and  
also the time period, $T$, of the dynamo for models with and without 
fluxes. In both panels it is clear that for $\Rm$ above
$\sim10^3$, ${\alpha_{\rm M}}$ and $T$ reach a saturated value.

\begin{figure}
\includegraphics[width=\columnwidth]{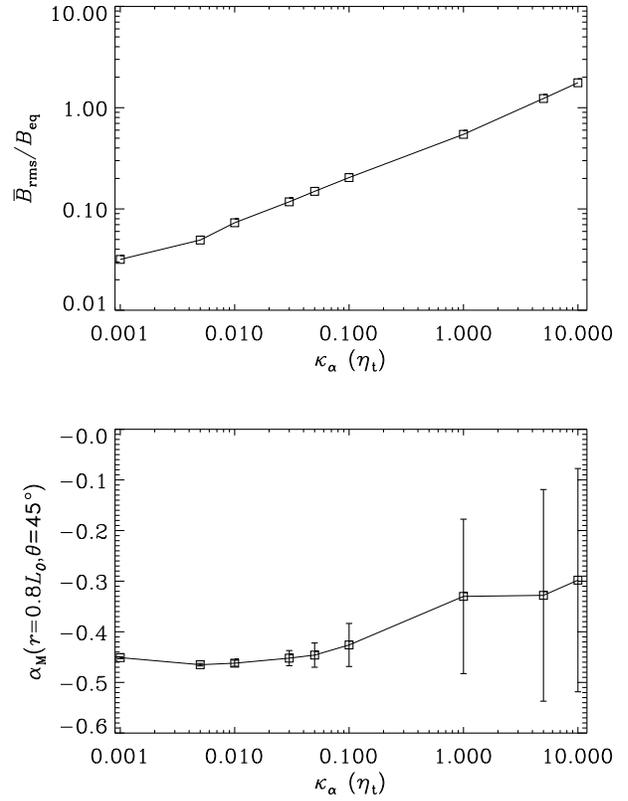}
\caption{Top, final amplitude of the rms mean magnetic field for different
  values of $\kappa$, in this case $R_{m}=10^7$. Bottom, final
  amplitude of $\alpha_{\rm M}$ at $r=0.8L_0$ and
  $\theta=45^{\circ}$. The error lines indicate the maximum and
  minimum amplitude in the oscillations of $\alpha_{\rm M}$ at this
  point.} 
\label{fig.bkappa}
\end{figure}

Besides its dependence on $\Rm$, the evolution of
$\alpha_{\rm M}$ depends also on $\kappa_{\alpha}$. For models with
$\kappa_{\alpha}\ll\eta_{\rm t}$, the evolution of 
$\alpha_{\rm M}$ relies on $R_{\rm m}$, but for
$\kappa_{\alpha}\ge0.1\etat$, the dissipation time of
$\alpha_{\rm M}$ becomes comparable to, or even shorter, than the
period of the dynamo cycle. This results in $\alpha_{\rm M}$ 
becoming oscillatory, as shown in the bottom panel of
Fig. \ref{fig.bkappa}. The amplitude and the period of these
oscillations depend on the value of $\kappa_{\alpha}$.

In the top panel of Fig. \ref{fig.bkappa} we show the final value of
the averaged mean magnetic field  as a function of
$\kappa_{\alpha}$. We observe that for 
$\kappa_{\alpha}$ in the range $(0.1$--$1)\,\etat$, the value of
$\overline{B}_{\rm rms}$ remains between 20\% and 60\% of the
equipartition, a value similar to the one obtained  in the
simulations using algebraic $\alpha$ quenching (Section~\ref{sec.4.1},
Fig.~\ref{fig.bif}). For  $\kappa_{\alpha}>\etat$, 
super-equipartition values of the 
magnetic field may be reached. This is because larger values of
$\kappa_{\alpha}$ result in oscillations of $\alpha_{\rm M}$ with larger 
amplitude, such $\alpha_{\rm M}$ may locally change its
sign, increasing the value of the total $\alpha$ in each
hemisphere and thereby enhancing the dynamo action. 
Such high values of the diffusion of the magnetic helicity are
unlikely in nature. 
\begin{figure}
\includegraphics[width=\columnwidth]{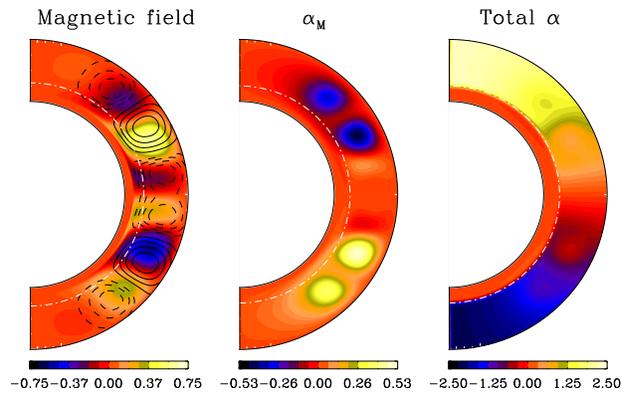}\\
\caption{The same than Fig. \ref{snaps.no.flux} but for a diffusive
  flux, with $\kappa=\eta$. The snapshot corresponds to
  $t=3.0$ ($L_0/\etat$).}
\label{snaps.flux}
\end{figure}

\subsection{The Vishniac-Cho flux}

Our next step is to explore the magnetic helicity flux proposed by 
\cite{vch01} in the form given by equation (\ref{eq.vcf}).
For the moment we set $\kappa_{\alpha}=0$.  In a
previous study on the effects of the 
VC flux in a MFD model in Cartesian coordinates, 
\cite{bs05c} found that there exist a critical value for the parameter
$C_{\rm VC}$ above which there is a runaway growth of the magnetic field
that can only be stopped using an additional algebraic quenching similar
to the one used in Section~\ref{sec.4.1}. 
They found that this critical value, $C_{\rm VC}*$, diminishes
with increasing the amount of shear. 
Since we have used a strong shear ($C_{\Omega}=-10^4$) we use
nominal values of $C_{\rm VC}=10^{-3}$, but without any algebraic
quenching.  

\begin{figure*}
\centering{
\includegraphics[width=115mm]{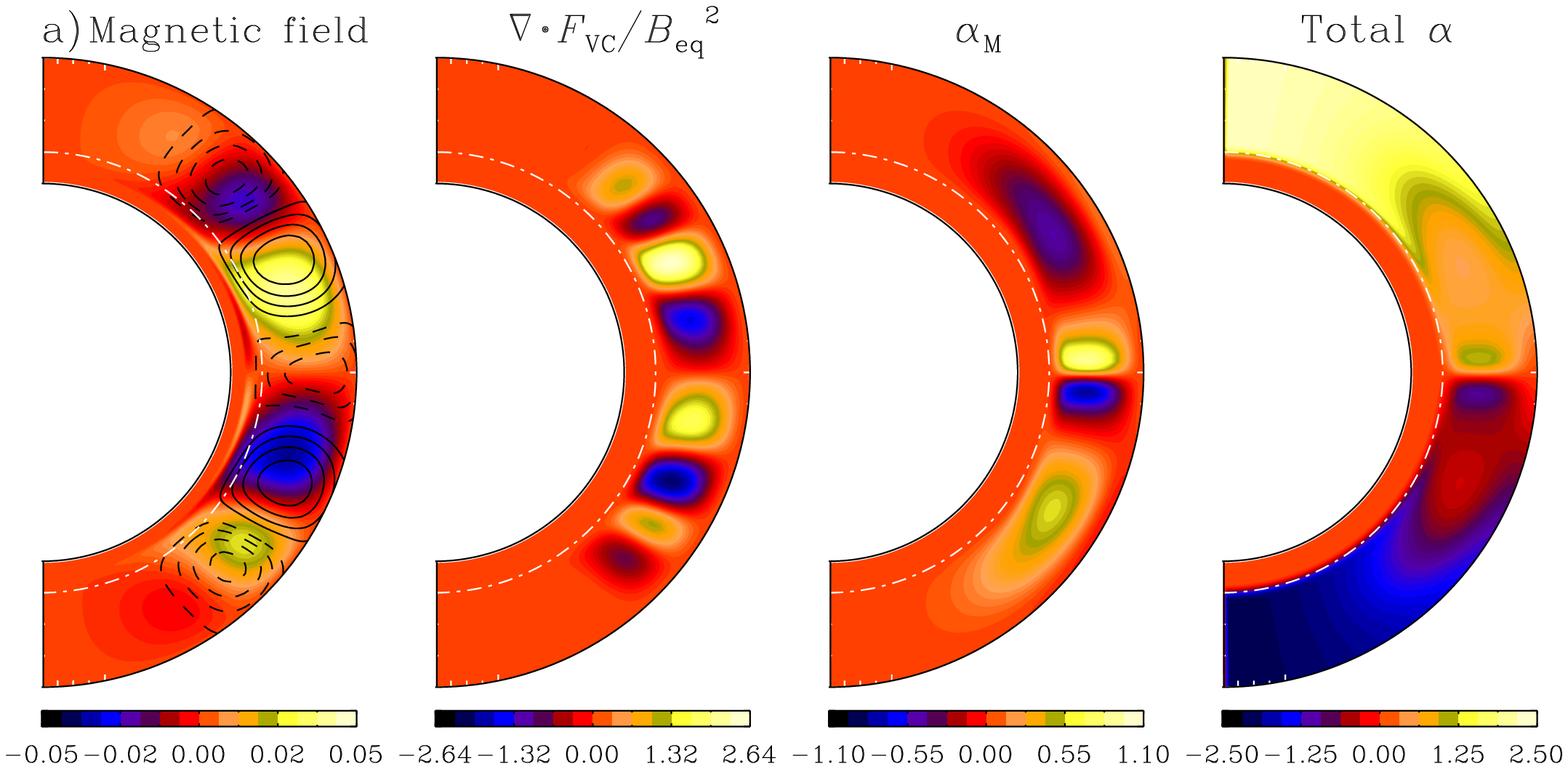}\\
\includegraphics[width=115mm]{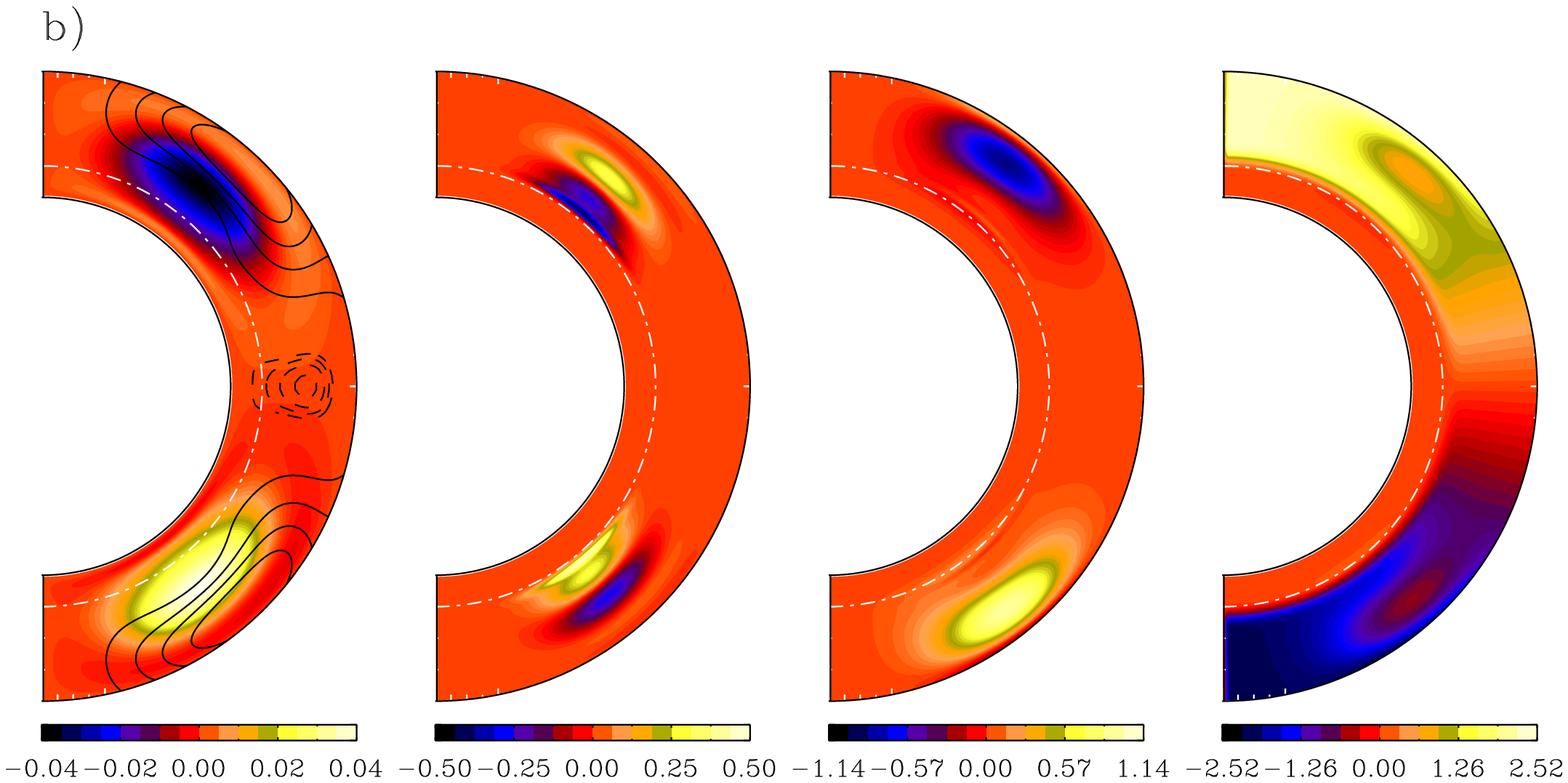}\
\includegraphics[width=115mm]{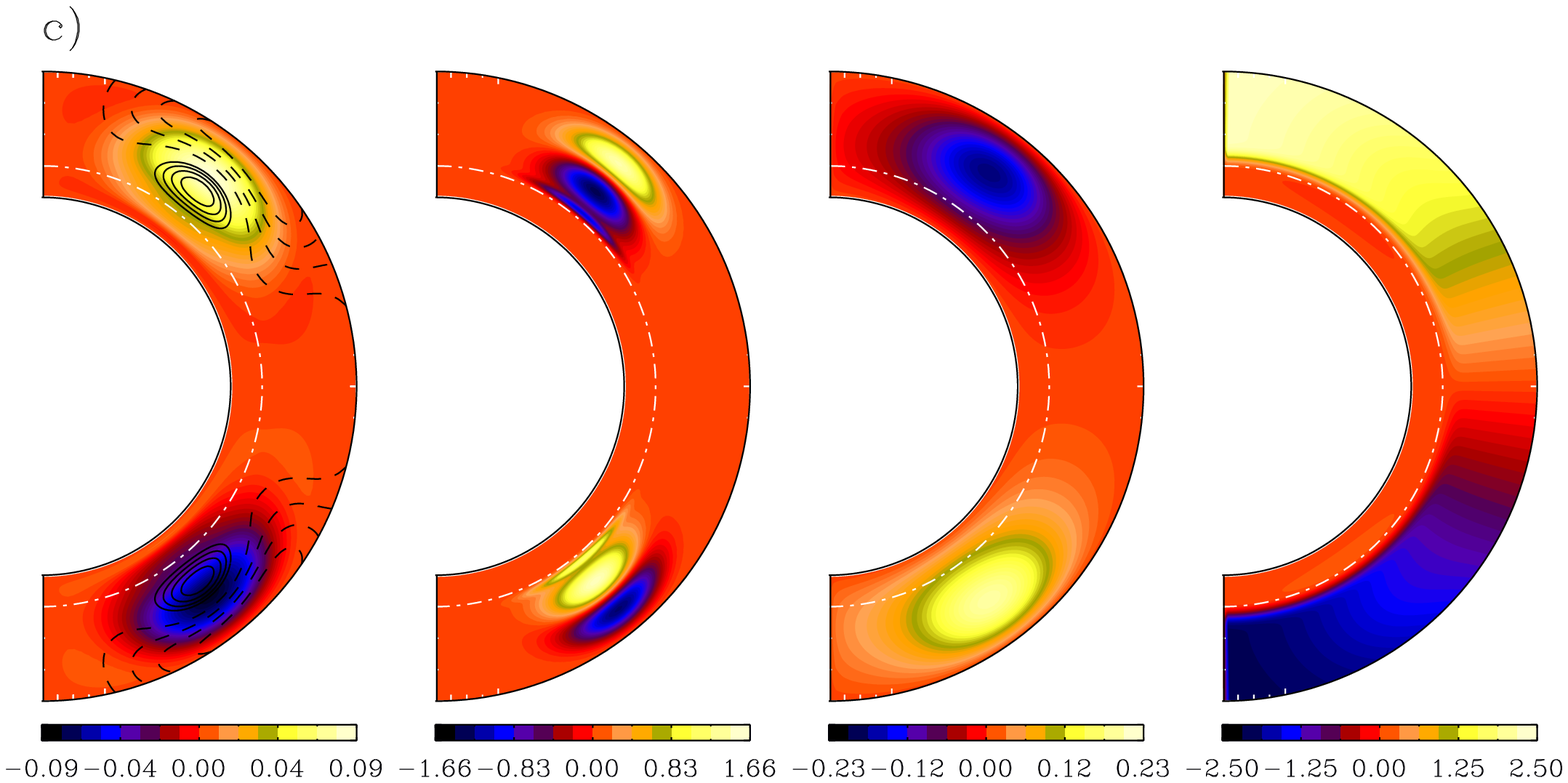}\\
\includegraphics[width=115mm]{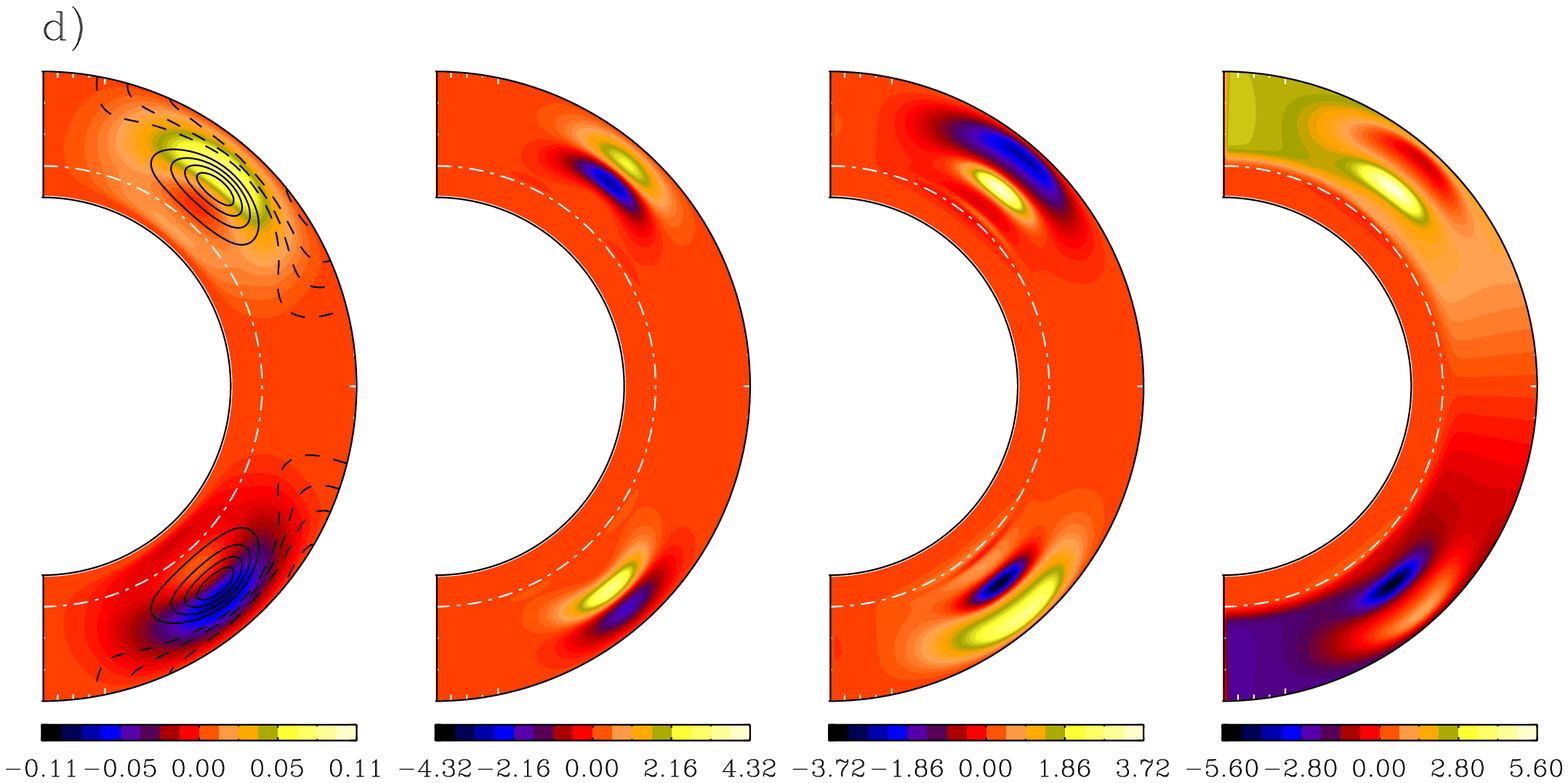}\\
\caption{Meridional snapshots of different models the in Table
  \ref{table1}: a) model VCa, b) model VC$_{\theta}$a, 
  c) model VCD$_{\theta}$ and d) model VC$_{\theta}$vf. The contours
  (colors and lines) for the magnetic 
  field have the same meaning than in Fig. \ref{snaps.no.flux}. In
  this plot we have include a new column with the value of the VC
  component in the $\alpha_{\rm M}$ equation, i.e.,
  $\bm\nabla\cdot\bm{\overline{\cal F}}_{\rm VC}/B^2_{\rm eq}$.
  All the snapshots corresponds to the relaxed state of evolution.}   
\label{snaps.vc.flux}}
\end{figure*}

\begin{figure}
\includegraphics[width=\columnwidth]{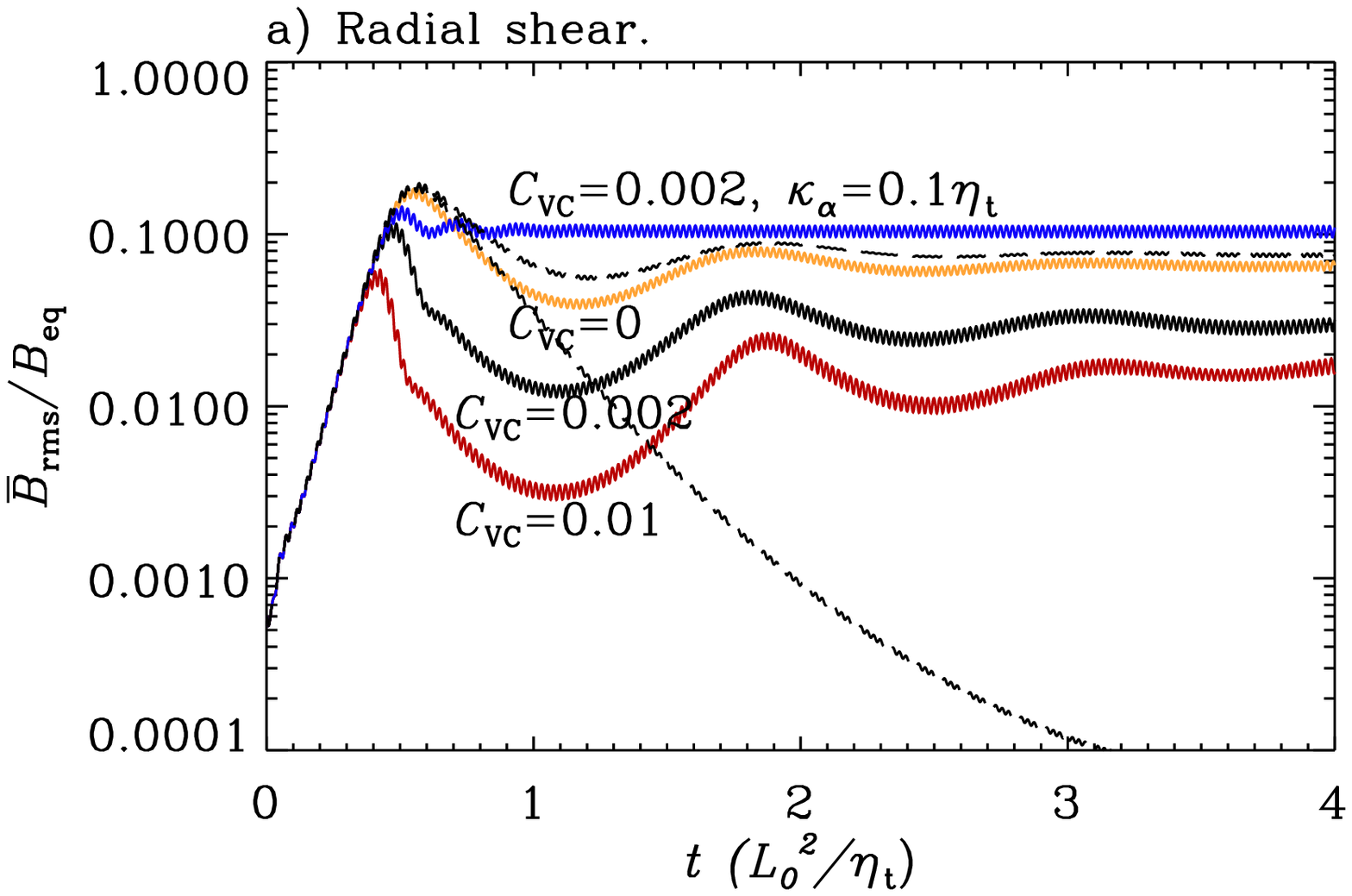}
\includegraphics[width=\columnwidth]{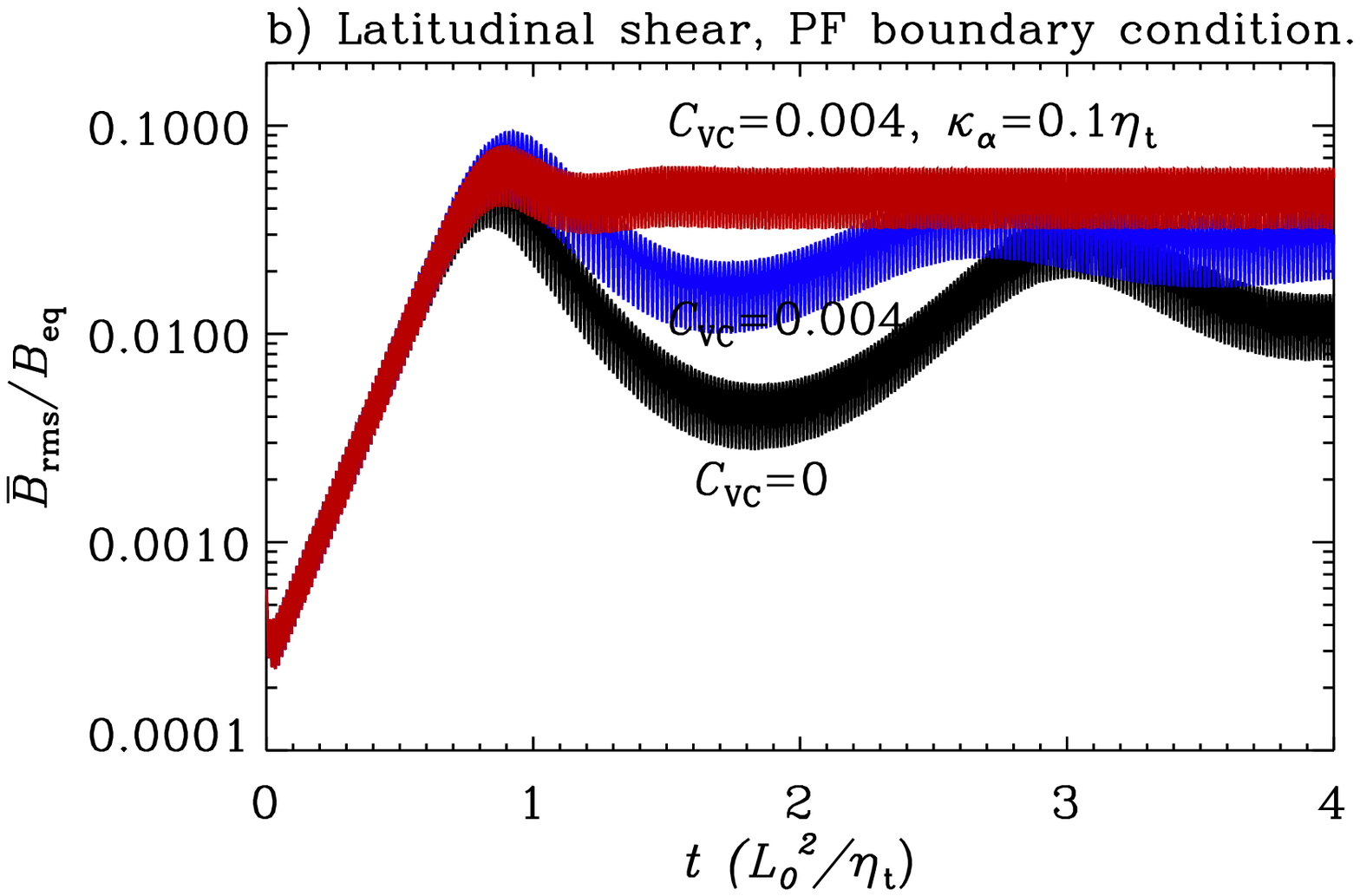}
\includegraphics[width=\columnwidth]{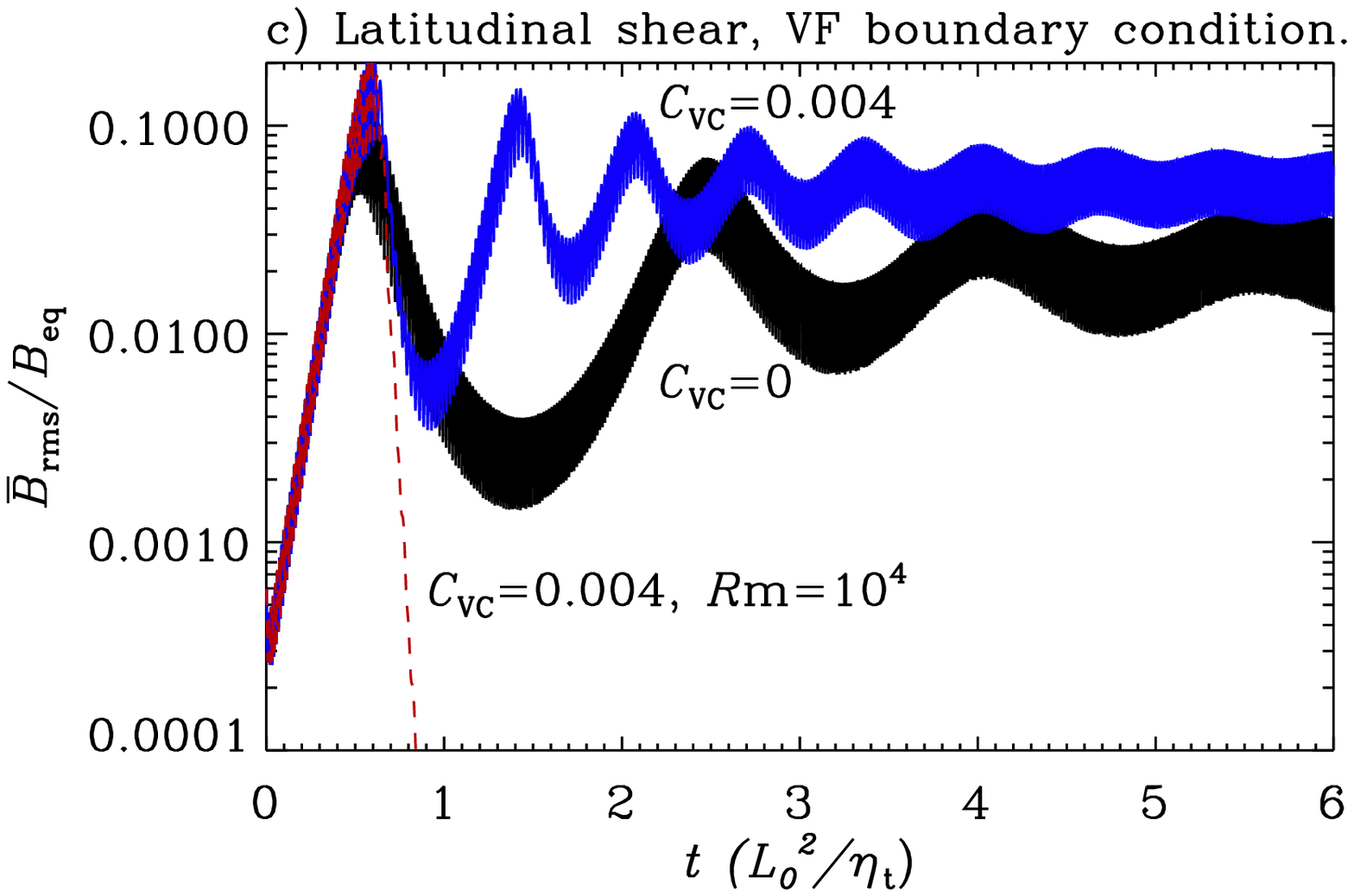}
\caption{Time evolution of the averaged mean magnetic field for
  different values of $C_{\rm VC}$: a) Radial shear, b) latitudinal shear
  with potential field boundary conditions and c) latitudinal shear
  with vertical field boundary conditions.
  The width of the different bands reflects the range over which
  the magnetic field varies during one cycle.
  Note that the cycle period is short compared with the resistive
  time scale on which the magnetic field reaches its final saturation.
  If not indicated, in all
  models $R_{\rm m}=10^3$. The two dashed lines in the panel a)
  corresponds to $C_{vc}=-0.002$ for $\Rm=10^3$ and
  $\Rm=10^4$.} 
\label{fig.en.vc.flux}
\end{figure}

\begin{figure}
\centering{
\includegraphics[width=\columnwidth]{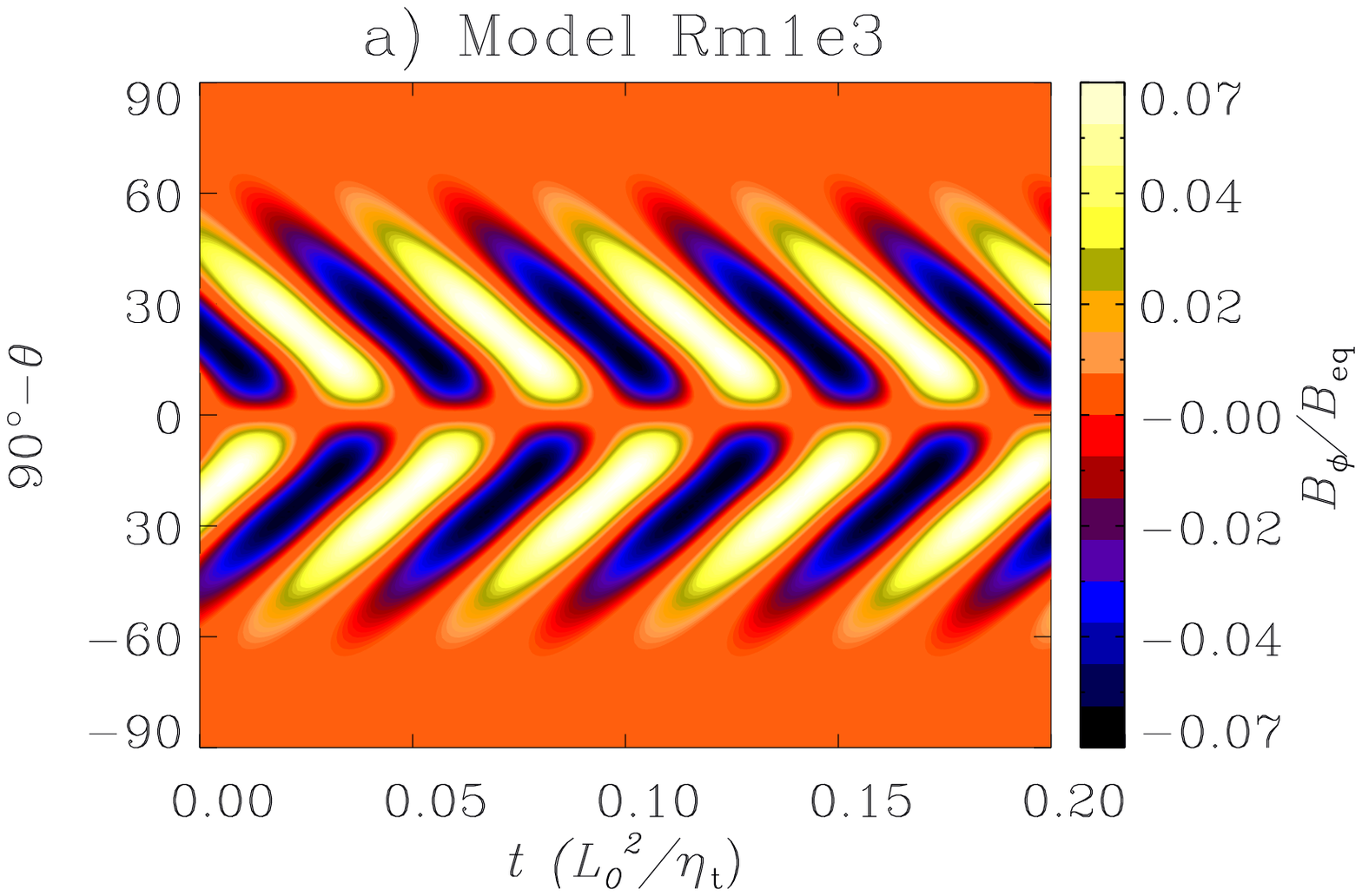}
\includegraphics[width=\columnwidth]{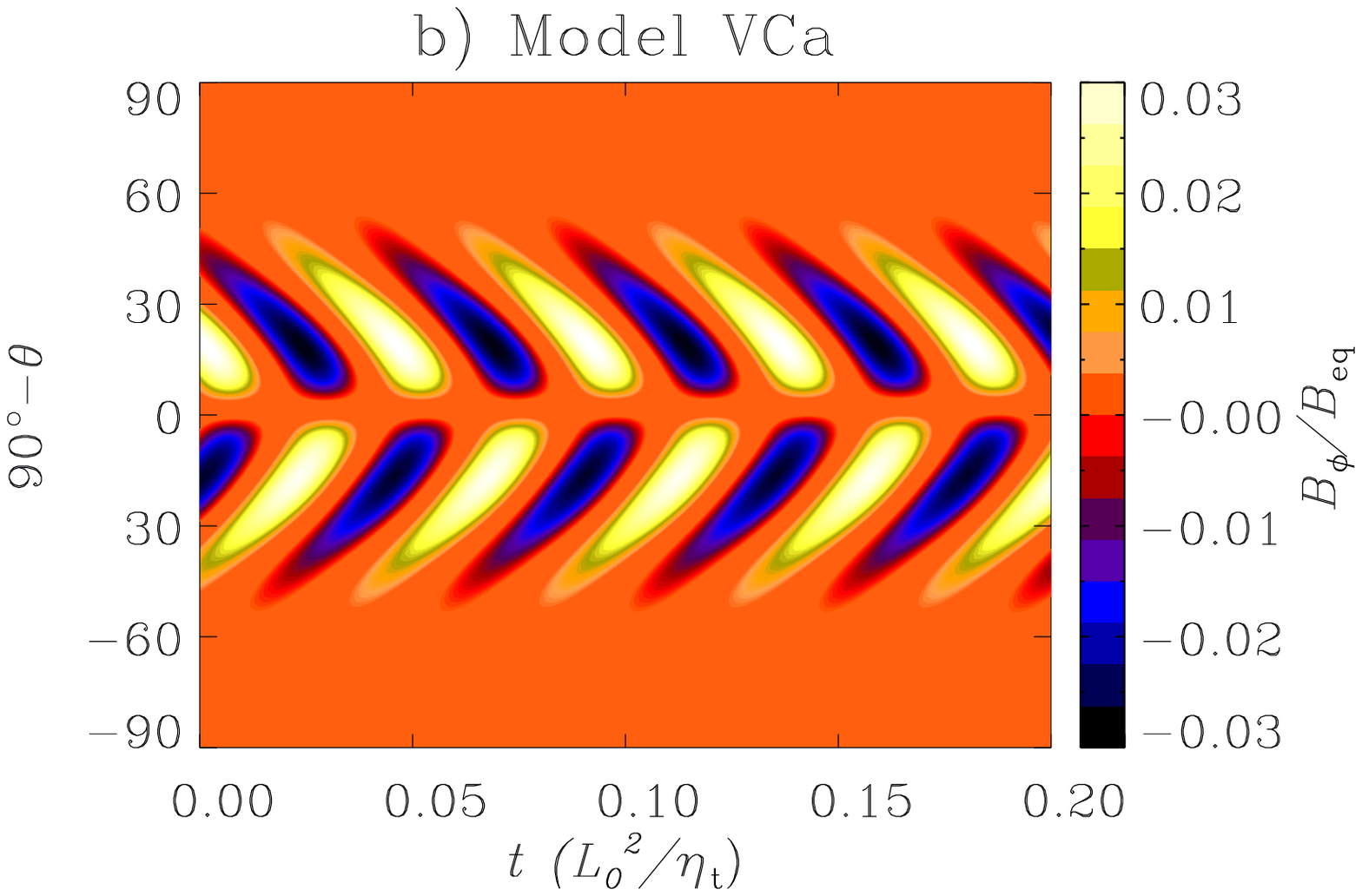}}
\caption{Butterfly diagrams of toroidal field for runs without
magnetic helicity flux (a) and with VC flux (b) for $R_{\rm m}=10^3$.
Note the stronger concentration of magnetic field at lower latitudes
in the presence of VC flux.}
\label{but.vc.flux}
\end{figure}

The term $\bm{\nabla}\cdot\overline{\bm{{\cal F}}}_{\rm VC}$ develops a
multi-lobed pattern which travels in the same direction as the dynamo
wave, this confirms that the VC flux 
follows the lines of iso-rotation. 
From equation (\ref{eq.vcf}), we see that the VC flux is
proportional to the magnetic energy density.
In the present case, with
$C_{\Omega} \gg C_{\alpha}$, the spatial distribution of
$\bm\nabla\cdot\bm{\overline{\cal F}}_{\rm VC}/B^2_{\rm eq}$ 
is dominated by the terms involving $B_{\phi}^2$ in  equations.~\ref{eq.vcfr}-\ref{eq.vcfp} (this may be inferred 
from the left hand panels of Fig.~\ref{snaps.vc.flux}a). 
This results in a new distribution 
of $\alpha_{\rm M}$, with concentrated regions of positive
(negative) sign at low latitudes in the northern (southern)
hemisphere, and a broad region of negative (positive) sign in
latitudes between $20^{\circ}$ and  $60^{\circ}$ latitude (see
middle panels of Fig. \ref{snaps.vc.flux}). Surprisingly we find that 
the general effect of this flux is to decrease the 
final amplitude of the magnetic field with respect to the case without
any fluxes as can be seen in Fig.~\ref{fig.en.vc.flux}. 
Note that we have until now used only the potential field
boundary condition for the poloidal field.
When we consider both diffusive as well as
VC fluxes, with $\kappa_{\alpha}=0.1\etat$ and
$C_{\rm VC}=10^{-3}$, we obtain a magnetic field of slightly larger 
amplitude compared to the case with only the diffusive flux (compare
the value of $\overline{B}_{\rm rms}$ in Runs~DRm1e3 and VCD in
Table~\ref{table1}). However we may say from the butterfly
diagram of Fig.~\ref{but.vc.flux} that the toroidal magnetic field
appears to be more concentrated at lower latitudes, where the sign of
$\alpha_{\rm M}$ is same as that of $\alpha_{\rm K}$. 

With negative values of $C_{\rm VC}$, it was found that the
resulting profile of $\alpha_{\rm M}$ is only weakly modified from cases
without fluxes, though its value is reduced marginally such that
the final amplitude of $\overline{B}_{\rm rms}$ is slightly
larger. But even this contribution does not help in alleviating
catastrophic quenching in models with large $R_{\rm m}$ (see
Fig.~\ref{fig.en.vc.flux}).  

Since VC fluxes transport helicity along lines of constant shear, 
it may be expected that they are more 
important in models with latitudinal shear, since in this case the
magnetic helicity flux can travel either towards the bottom or the
top boundaries, from where magnetic helicity can be expelled. For
testing this possibility, we turn off the radial shear 
profile and consider a purely latitudinal solar-like differential
rotation: 
\begin{equation}
\Omega(r,\theta) = C_{\Omega}\left(\frac{\etat}{\Omega_{\rm eq}L_0^2}
\right) \Theta(r,r_2,w_1)
(\Omega_s(\theta)-\Omega_c) \; ,
\end{equation}
\noindent 
where $\Omega_{\rm eq}/2\pi=460.7$ nHz is the
angular velocity at the equator, and
$\Omega_s(\theta)=\Omega_{\rm eq}+a_2\cos^2 \theta+a_4\cos^4 \theta$
gives the latitudinal profile, with
$a_2/2\pi=-62.9$ nHz and $a_4/2\pi=-67.13$ nHz.
 
In order for the dynamo to be slightly
supercritical, as in the previous cases, we consider
$C_{\Omega}=5\times 10^4$. 
This dynamo solution corresponds to a dynamo wave produced at mid
latitudes ($\sim45^{\circ}$) that travels upwards (since
$C_{\Omega}$ now is positive). 
As in the previous cases with radial shear, the distribution of 
$\bm\nabla\cdot\bm{\overline{\cal F}}_{\rm VC}/B^2_{\rm eq}$ is
similar to that of the divergence of magnetic energy density
(left hand panels of Fig.~\ref{snaps.vc.flux} b,c and d).
If no fluxes are considered, the final amplitude of the 
mean magnetic field is $\sim 0.03$\% of the equipartition value. In
presence of VC fluxes, starting with $C_{\rm VC}=10^{-3}$ for a
model with $R_{\rm m}=10^3$, we notice that the final 
magnetic field is twice as large as in the case with $C_{\rm VC}=0$. 

Our model becomes numerically unstable beyond $C_{\rm VC}=10^{-2}$ 
due to appearance of concentrated regions of strong $\alpha_{\rm M}$. 
When VC and diffusive fluxes are considered simultaneously,
with $C_{\rm VC}=10^{-3}$ and $\kappa_{\alpha}=0.1\etat$, the
relaxed value of $\overline{B}_{\rm rms}$ is only slightly
below the value reached at the end of the kinematic phase
(Fig.~\ref{fig.en.vc.flux}b). In this case $\alpha_{\rm M}$ spreads
out in the convection zone, as shown in Fig. \ref{snaps.vc.flux}c,
indicating that the effects of the VC flux are not important when
compared with the diffusive flux.

We repeated the calculation by considering the vertical field (VF)
boundary condition, $\partial (r B_{\theta})/\partial \theta=0$,
for the top boundary, instead of the potential field (PF) condition
used throughout the rest of this work.
Furthermore, in the models with VF conditions the presence of the VC flux
leads to an increase of $B_{\rm sat}$ by a factor of $\sim 2$ compared
to the case without VC flux (see Fig.~\ref{fig.en.vc.flux}c).
It may be noted that $\alpha_{\rm M}$ shows 
regions of both positive and negative signs in each
hemisphere (see Fig.~\ref{snaps.vc.flux}d).
Thus, the total $\alpha$ effect is increased locally to values well
above the kinematic one. This implies that in the region
around $\pm 45^\circ$ the dynamo action is driven by the magnetic
$\alpha$ effect. 
A similar secondary dynamo is found to be working 
for a different distribution of shear and $\alpha_{\rm K}$
\citep{cgb10}.
As with PF boundary condition, large
values of $C_{\rm VC}$ result in a numerical instability of the
magnetic field in the simulation with VF.

The main result of this section is that the VC flux does not
alleviate catastrophic quenching of the 
dynamo for large values of $\Rm$ (see the dashed lines in
Fig.~\ref{fig.en.vc.flux} a and c).
The reason for this may be related to the fact that the radial flux
has components that are either proportional to $B_\theta$ or to
$B_\phi$ (equation \ref{eq.vcfr}).
As $B_\phi$ vanishes on the top boundary, and $B_\theta$ is small,
the VC flux is not able to dispose of $\alpha_{\rm M}$ across the
boundary. This might change if diffusive fluxes became important near
the top or if a different boundary condition on $B$ were applied.

\section{Conclusions}
\label{sec.conc}
We have developed $\alpha\Omega$ dynamo models in spherical geometry
with relatively simple profiles of $\alpha_{\rm K}$ and shear
($\partial \Omega/\partial r$ and $\partial \Omega/\partial
\theta$). We choose potential field (also vertical field in some 
  cases) and perfect conductor boundary conditions for the top and
bottom boundaries, respectively.  We estimate the critical dynamo
number by fixing $C_{\Omega}=-10^4$ and varying $C_{\alpha}$ while
using  algebraic quenching.

Using a dynamo number, $C_{\Omega}C_{\alpha}$, that is slightly
super-critical, we solve  
the induction equations for $B$ and $A$ 
together with an equation for the
dynamical evolution of the magnetic $\alpha$ effect or $\alpha_{\rm M}$. We
find that for positive (negative) values of $C_{\alpha}$ in the northern
(southern) hemisphere, $\alpha_{\rm M}$ is mainly negative (positive), with
narrow fractions of opposite sign in regions where $\alpha_{\rm K}$ or
$\overline{B}$ 
are equal to zero. 

We find that the kinematic phase is independent of $\Rm$. However for
$\Rm>10^2$ there exists a phase of relaxation post saturation in which
the averaged magnetic field oscillates about a certain mean.
The larger the $\Rm$, the more
pronounced are the damped oscillations and the
longer is the relaxation time (Fig. \ref{fig.no.flux}). The final
value of the magnetic energy obeys a $\Rm^{-1}$ dependency ($R_{\rm
  m}^{-0.5}$ for magnetic field, Fig. \ref{fig.brms}), which is in
agreement with earlier work \citep{bs05c,bcc09}. 

We argue that including equation (\ref{eq.am}) in MFD models is
appropriate  
for describing the quenching of the magnetic field in the dynamo
process. 
Since we observe large-scale magnetic fields at high magnetic Reynolds
numbers in astrophysical objects, there must exist a mechanism to
prevent the magnetic field from catastrophic quenching.

We have studied the role that diffusive and VC fluxes may play in
this sense. Their contribution may be summarized as follows: 
\begin{enumerate}
\item
In the presence of diffusive fluxes, $\alpha_{\rm M}$ has only one sign in 
each hemisphere (negative in the northern hemisphere and positive in
southern) and is evenly distributed across the dynamo region
(Fig. \ref{snaps.flux}).  
\item
For $\Rm<10^2$ the mean values of $\alpha_{\rm M}$ are similar to models
without diffusive fluxes, whereas for $\Rm\ge10^2$, $\alpha_{\rm
  M}$ has smaller values that seem to be independent of $R_{\rm m}$
(see Fig. \ref{fig.brms}, middle).
\item
Even a very low diffusion coefficient, e.g.\ $\kappa_{\alpha}=0.001\etat$,
causes $\overline{B}_{\rm rms}$ to depart from the $R_{\rm
  m}^{-0.5}$ tendency and converge to a constant value which is then around
$5$\% of the equipartition value for large values of $\Rm$,
but below the value of $10^7$ used in
this study (dashed line in Fig. \ref{fig.brms}, top). 
\item
Larger values of $\kappa_{\alpha}$ result in larger final field
strengths. 
\item
In models with only radial shear the Vishniac-Cho flux contributes to
$\alpha_{\rm M}$ with a component that travels in the same direction as
the dynamo wave. This produces a different radial and latitudinal
distribution of the magnetic $\alpha$ effect that also affects the  
distribution of the magnetic fields.
However, it does not help in alleviating the quenching at high $R_{\rm
  m}$. On the contrary, the larger the coefficient $C_{\rm VC}$, the
smaller is the resultant magnetic field.
\item
In models with only latitudinal shear the VC flux travels 
radially outward but it remains concentrated at the center of the
dynamo region. 
In a given hemisphere the resultant distribution of $\alpha_{\rm M}$
has both positive and negative signs. The part of $\alpha_{\rm M}$
that has the same sign as $\alpha_{\rm K}$ enhances dynamo action.
This effect is more evident in models with vertical field boundary
conditions (Figs. \ref{snaps.vc.flux}b-d). 
\item
In models with vacuum and vertical field boundary conditions and
$R_{\rm m}=10^3$,  
the VC flux increases the final value of the magnetic field by a 
factor of two compared to the case without any fluxes.
\item 
The magnetic field in models with $R_{\rm m}\geq 10^4$ and with 
non-zero VC flux decays after the kinematic phase since the total
$\alpha$ effect becomes subcritical (see dashed lines in
Fig. \ref{fig.en.vc.flux} a and c). 
\item
Larger values of $C_{\rm VC}$ produce narrow bands of 
$\alpha_{\rm M}$ 
which drives intense dynamo action in these regions. 
This positive feedback between the magnetic field and $\alpha_{\rm M}$
causes the simulation to become numerically unstable
in the absence of any other quenching effect.
\end{enumerate}

From the above results it is clear that diffusive fluxes are much more
important in alleviating catastrophic quenching when compared to the
Vishniac \& Cho fluxes (in the form of equation \ref{eq.vcf}) for a
large range of $\Rm$. This is somehow intriguing since it is known
from DNS that shear in domains with open boundaries does indeed help
in alleviating the catastrophic quenching. 
It may be understood as a result of the large value of $C_{\Omega}$ 
compared with $C_{\alpha}$ and also to the top boundary condition
for the azimuthal magnetic field \citep{B05,kapyla08}.

The results presented above  
indicate that considerable work is still necessary in order to
understand the role of larger-scale shear in transporting and shedding
small-scale magnetic helicity from the domain. 

In snapshots of the meridional plane as well as in butterfly
diagrams we notice that the diffusive fluxes do not significantly
modify the morphology and the distribution of the magnetic field when
compared with cases without fluxes or even with simulations with
algebraic $\alpha$ quenching. On the other hand, for models 
with VC flux the distribution of $\alpha_{\rm M}$ becomes
different and so does the magnetic field. 
This is clear from the butterfly diagram shown in
Fig.~\ref{but.vc.flux}b, which exhibits a  
magnetic field confined to equatorial latitudes reminiscent of
the observed butterfly diagram of the solar cycle. Even though this
result corresponds to a simplified model, it illustrates the
importance of considering the dynamical $\alpha$ quenching mechanism 
for modeling the solar dynamo. 
Similar changes in the distribution of $\alpha_{\rm M}$ and
$\bm{\overline{B}}$ are expected to happen when advection terms are
included in the governing equations.

In the simulations presented here, $\Omega$ and $\alpha$ effects
are present in the same layers. An interesting question is whether the 
quenching of the dynamo is catastrophic when both layers are
segregated, as in the Parker's interface dynamo or the 
flux-transport dynamo models. We address this question in detail in
two companion papers \citep{cbg10,cgb10}.

We should notice that the back reaction of the magnetic field
affects not only the $\alpha$ effect, but also
the other dynamo coefficients, including the turbulent diffusivity. 
Contrary to quenching of $\alpha$, the quenching of $\etat$ may be
considered through an algebraic quenching function \citep[see
  e.g.][]{ybr03,kb09}. 
\cite{gddp09} have shown that in a flux-transport model these effects
could affect properties of 
the models such as the final magnetic field strength and its distribution
in radius and latitude. 
We leave the study of models with simultaneous
dynamical $\alpha$ and $\eta$ quenchings for a future paper.
Solar-like profiles of differential rotation and meridional circulation
along with dynamical $\alpha$ quenching will also be considered in a
forthcoming paper.

\section*{Acknowledgments}
This work started during the NORDITA program solar and stellar dynamos
and cycles and is supported by the European Research Council under the
AstroDyn research project 227952.

\newcommand{\yjfm}[3]{ #1, {J.\ Fluid Mech.,} {#2}, #3}

\label{lastpage}
\end{document}